\documentclass[fleqn,usenatbib,useAMS]{mnras}

\usepackage{newtxtext,newtxmath}

\usepackage[T1]{fontenc}
\usepackage{ae,aecompl}
 
\usepackage{graphicx}	
\usepackage{amsmath}	
\usepackage{amssymb}	
\usepackage{longtable}
\usepackage{natbib}
\usepackage{multirow}
\usepackage{lscape} 
\usepackage{tabularx}

\newcommand{\orcid}[1]{\href{https://orcid.org/#1}{\includegraphics[width=10pt]{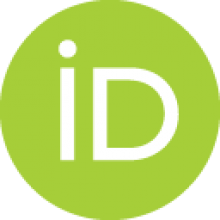}}}

\newcommand{\lsim}{\raisebox{-.5ex}{$\,\stackrel{\textstyle <}{\sim}\,$}}
\newcommand{\gsim}{\raisebox{-.5ex}{$\,\stackrel{\textstyle >}{\sim}\,$}} 
\title[Orbital classification of N-rich stars]{Dynamical Orbital classification of selected N-rich stars with \textit{Gaia} DR2 astrometry}

\author[Fern\'andez-Trincado et al.]{
	Jos\'{e} G. Fern\'{a}ndez-Trincado$^{1}$\thanks{E-mail: jose.fernandez@uda.cl}\orcid{0000-0003-3526-5052},
    Leonardo Chaves-Velasquez$^{2,3}$\orcid{0000-0002-9677-1015},
	Angeles P\'erez-Villegas$^{4}$\orcid{0000-0002-5974-3998},
	\newauthor
	Katherine Vieira$^{1}$\orcid{0000-0001-5598-8720},	
	Edmundo Moreno$^{5}$,
	Mario Ortigoza-Urdaneta$^{1}$\orcid{0000-0003-4092-7655}
	\newauthor
	and
	Luis Vega-Neme$^{6,7}$
	\\
	$^{1}$Instituto de Astronom\'ia y Ciencias Planetarias, Universidad de Atacama, Copayapu 485, Copiap\'o, Chile.\\
    $^{2}$University of Nari\~no Observatory, Universidad de Nari\~no, Sede VIIS, Avenida Panamericana, Pasto, Nari\~no, Colombia.\\
    $^{3}$Departamento de F\'isica de la Universidad de Nari\~no, Torobajo Calle 18 Carrera 50, Pasto, Nari\~no, Colombia.\\
	$^{4}$Universidade de S\~ao Paulo, IAG, Rua do Mat\~ao 1226, Cidade Universit\'aria, S\~ao Paulo 05508-900, Brazil.\\
	$^{5}$Instituto de Astronom\'ia, Universidad Nacional Aut\'onoma de M\'exico, Apdo. Postal 70264, M\'exico D.F., 04510, M\'exico.\\
	$^{6}$Universidad Nacional de C\'ordoba, Observatorio Astron\'omico de C\'ordoba, C\'ordoba, Argentina.\\
	$^{7}$CONICET, Instituto de Astronom\'ia Te\'orica y Experimental, C\'ordoba, Argentina.\\
}

\date{Accepted XXX. Received YYY; in original form ZZZ}

\pubyear{2020}

\begin{document}
\label{firstpage}
\pagerange{\pageref{firstpage}--\pageref{lastpage}}
\maketitle

\begin{abstract}
We have used the galaxy modeling algorithm \texttt{GravPot16}, to explore the more probable orbital elements of a sample of 64 selected N-rich stars across the Milky Way. Using the newly measured proper motions from \texttt{Gaia} DR2 with existing line-of-sight velocities from APOGEE-2 survey and spectrophotometric distance estimations from the \texttt{StarHorse}. We adopted a set of high-resolution particle simulations evolved in the same steady-state Galactic potential model with a bar, in order to identify the groups of N-rich stars that have a high probability of belonging to the bulge/bar, disk, and stellar halo component. We find that the vast majority of the N-rich stars, show typically maximum height from the Galactic plane below 3 kpc, and develop rather eccentric orbits (\textit{e}$>$0.5), which means these stars appear to have bulge/bar-like and/or halo-like orbits. We also show that $\sim66$\% of the selected N-rich stars currently lives in the inner Galaxy inside the corotation radius (C.R.), whilst $\sim14$\% of the N-rich star resides in halo-like orbits. Among the N-rich in the inner Galaxy, $\sim27\%$ of them share orbital properties in the boundary between bulge/bar and disk, depending on the bar pattern speeds. Our dynamical analysis also indicates that some of the N-rich are likely halo interlopers and therefore suggest that halo contamination is not insignificant within the bulge area.
\end{abstract}




\section{INTRODUCTION}
\label{section1}

	Large spectrocopic surveys like APOGEE \citep{Majewski2017}, capable to see through the dusty part of the Milky Way, together with other complementary surveys such as Gaia-ESO survey \citep[][among other]{Gilmore2012, Randich2013}, have clearly revealed that our Galaxy hosts a significant population of giant stars with strong $^{12}$C$^{14}$N features in their atmospheres residing preferentially in the bulge and stellar halo of the Milky Way \citep[e.g.,][]{Martell2016, Pereira2017, Fernandez-Trincado2016b, Fernandez-Trincado2017a, Fernandez-Trincado2017b, Fernandez-Trincado2018, Reis2018, Fernandez-Trincado2019b, Fernandez-Trincado2019c, Koch2019}.  

	APOGEE high-resolution spectra (\textit{R}$\approx$ 22,500) on the near-IR \textit{H}-band ($\lambda \sim$ 1.5--1.7 $\mu$m) have shown that many of these stars are often typified by large Nitrogen-enhancements ([N/Fe]$\gsim+0.5$, hereafter N-rich) simultaneous with depletion in carbon ([C/Fe]$\lsim+0.15$), identified most obviously by their $^{12}$C$^{16}$O-band, enhanced $^{12}$C$^{14}$N-band features \citep[see, e.g.,][]{Altmann2005, Martell2010, Martell2011, Lind2015, Martell2016, Fernandez-Trincado2019b}. Other abundances of elements involved in proton-capture reactions, i.e., Mg and Al, often show structured coherent patterns. Futhermore, the vast majority of the N-rich stars investigated so far possess clear chemical abundance patterns in their light elements, similar to the patterns observed in \textit{second-generation} globular cluster stars (enhanced N and Al, and depleted Mg, C and O abundances, with respect to field stars at the same metallicity). 
	
	The current best interpretation is that such an unusual field stars are former globular cluster stars \citep{Khoperskov2018, Savino2019}, and as such, play an important role in deciphering the early history of the Milky Way itself \citep{Martell2010, Carollo2013, Kunder2014, Lind2015, Recio-Blanco2017, Koppelman2018, Helmi2018, Ibata2019, Fernandez-Trincado2019b,  Fernandez-Trincado2019a,  Fernandez-Trincado2019c, Fernandez-Trincado2019d, Tang2019, Tang2020}. Their origin can also be ascribed to different exotic events, including external mechanisms such as binary mass-transfer channel and/or "in-situ" formation, and \citet{Fernandez-Trincado2019a} recently detected a Nitrogen enriched field star formed through the binary channel. These scenarios have recently been questioned \citep[see, e.g.,][]{Bekki2019}, suggesting that the vast majority of the N-rich stars in the Galactic stellar halo originate not from globular clusters (GCs) but have the same origin as those in the bulge. Currently, there is no real working explanation for the origin of the gamut of extreme light and heavy elements simultaneously intervening in the chemical composition of these newly identified anomalous field stars. One of the most striking results from the APOGEE survey was discovered by \citet{Fernandez-Trincado2017a}: a very exotic clump (7 out of 11 stars) of giant stars in the metal-poor tail ([Fe/H] $\lsim-0.7$) of the thick disc metallicity distribution, with significant Mg-underabundances and extreme enrichment in N and Al. A possible link to extragalactic globular clusters \citep{Pancino2017} was sugested, as none of the well studied Galactic globular clusters in our own Galaxy can reproduce this. The dynamical properties of such population remains unexplored to date.
	
	Beyond the intrinsic value of identifying chemically anomalous field stars throughout the Milky Way and understanding the related abundance phenomena, knowing their orbital parameters may offer a better insight into the origin of such stars. The newly discovered N-rich stars in the Milky Way \citep[e.g.,][]{Martell2016, Fernandez-Trincado2017a} have a number of essential parameters that remain unexplored. In an effort to remedy this deficiency, we conducted for the first time a dynamical orbital classification of such stars to predict their orbital path across the Milky Way as well as reveal their birthplace.
	
	In this work we take advantage of the accurate proper motions of the European Space Agency's Gaia mission Second Data Release (DR2) archive \citep{Lindegren2018,Arenou2018}, complemented along with radial velocities from APOGEE-2 \citep{Nidever2015} and spectro-photometric distances from \texttt{StarHorse} \citep{Queiroz2019}. They allow us an unprecedented combination of precision to fully resolve the space velocity and position vectors of the targeted stars in order to study for the first time their dynamical behavior across the Milky Way, in a non-axisymmetric Galactic model, like \texttt{GravPot16}\footnote{\url{https://gravpot.utinam.cnrs.fr/}}. 
	
	This paper is outlined as follows. In \S\ref{section2}, we describe the APOGEE and Gaia data set and a series of selection criteria. In \S\ref{section3}, we present the details of the Galactic model, followed by a discussion of the constraints on the free parameters of the model and the choices for the fiducial parameter values. We then move on to discuss the dynamical properties of selected N-rich stars in \S\ref{section4} and we summarize their dynamical orbital classification in \S\ref{section5}. Finally, \S\ref{section6} present our conclusions.

\begin{figure*}
	\begin{center}
		\includegraphics[width=170mm]{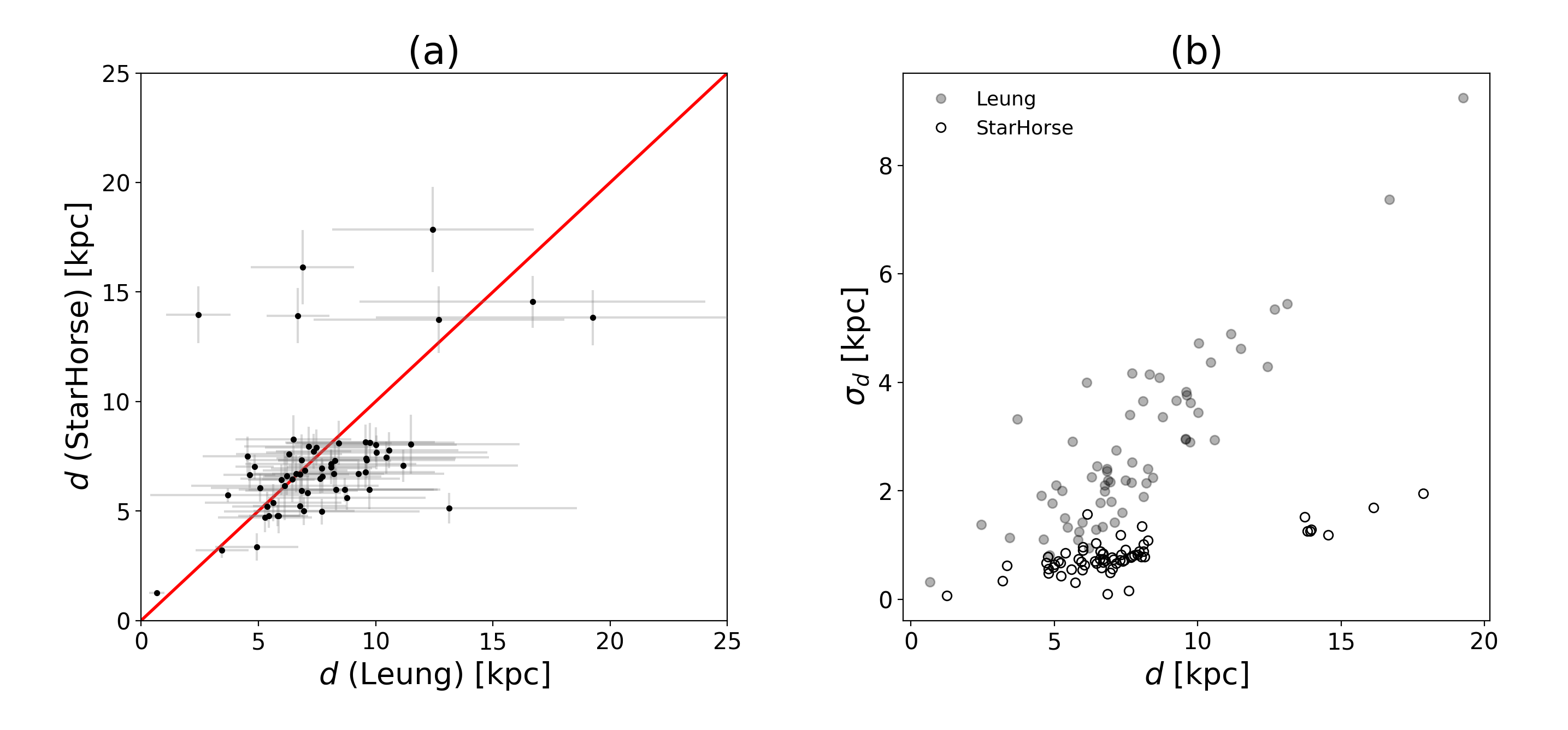}
		\caption{Comparison between our inferred distances from APOGEE DR16 \texttt{StarHorse} and the distances from \citet{Leung2019}. In panel (a) the red solid line marks the 1:1 relation. The panel (b) shows the uncertainty as a function of distances.}
		\label{Figure1}
	\end{center}
\end{figure*}

\section{THE OBSERVATIONAL SAMPLE OF STARS}
\label{section2}

The sample analysed in this work consist of N-rich stars located towards the bulge, disc and halo taken from \citet{Martell2016}, \citet{Fernandez-Trincado2016b}, \citet{Schiavon2017} and \citet{Fernandez-Trincado2017a}. These unusual field giant stars have been widely explored in the APOGEE survey \citep{Majewski2017}, through the \textit{H}-band ($\lambda \sim$ 1.5--1.7 $\mu$m) high-resolution (R$\sim$22,500) APOGEE spectra, obtained with the 300-fiber spectrograph installed on the 2.5m Telescope \citep{Gunn2006} at the Apache Point Observatory as part of the Sloan Digital Sky Survey IV \citep{Blanton2017}. The reduction of the APOGEE spectra, as well as the determination of radial velocities, atmospheric parameters and stellar abundances were carried out by the \texttt{ASPCAP} pipeline \citep[see][]{Nidever2015, Zamora2015, Holtzman2015, Jonsson2018, Holtzman2018}, using reduction scripts designed for the 16$^{th}$ data release of SDSS \citep[DR16,][]{Ahumada2020}. We refer the reader to \citet{Zasowski2013} and \citet{Zasowski2017} for full details regarding the targeting strategies of APOGEE and APOGEE-2.

\subsection{6-D datasets}

By cross-matching APOGEE-2 data to the \texttt{Gaia} DR2 \citep{gaiadr2}, we have found Gaia counterparts to most of the N-rich stars studied in \citet{Martell2016}, \citet{Fernandez-Trincado2016b}, \citet{Schiavon2017} and \citet{Fernandez-Trincado2017a}, with 5-D phase-space information. Only 7 of these stars have \texttt{Gaia} radial velocity information, with uncertainties on the order of 0.3--11 km s$^{-1}$. Our study uses APOGEE line-of-sight velocity because they are generally higher precision, in our sample typically $\lsim$ 2 km s$^{-1}$, compared to other catalogues. We select our \texttt{Gaia} DR2 sub-sample based mainly on the recommended \texttt{RUWE} (renormalized unit weight error) astrometric quality indicator. By selecting only those GAIA DR2 sources with \texttt{RUWE} $\lsim$ 1.40, we ensure to have astrometrically well-behaved sources \citep[see e.g.,][]{Lindegren2018}.

\subsection{Distances}
\label{distance}

We employed precise spectrophotometric distances (with median uncertainties of $\lsim$2 kpc) estimated with the Bayesian \texttt{StarHorse} code and published in an SDSS Value Added Catalogue\footnote{\url{https://www.sdss.org/dr16/data_access/value-added-catalogs/?vac_id=apogee-dr16-starhorse-distances-and-extinctions}}
\citep[see,][for more details]{Santiago2016, Queiroz2018, Queiroz2019}.
It combines atmospheric parameters (T$_{\rm eff}$, log \textit{g} and [M/H]) from the processed \texttt{ASPCAP} pipeline, multiband photometric information (APASS, 2MASS, and All- WISE) and the \texttt{Gaia} astrometric information when available along with their associated uncertainties, accounting for the global \texttt{Gaia} parallax zero-point shift of $-$0.029 mas \citep{Lindegren2018, Arenou2018}. The choice to use \texttt{StarHorse} spectrophotometric distances for the orbital integrations we will describe later, is based on the fact that for  distant  stars as those in the Galactic bulge and halo, parallax measurements can be uncertain ($>$ 22\% and extending all the way up to 1400\% uncertainty in the most extreme case). 

A total of 64 out of 76 sources have reliable \texttt{StarHorse} spectrophotometric distances and \texttt{RUWE}$<1.40$: 
	(\textit{i}) One (1) N-rich star, TYC 5619-109-1, from \citet{Fernandez-Trincado2016b}; 
	(\textit{ii}) Four (4) N-rich stars from \citet{Martell2016}; 
	(\textit{iii}) Nine (9) N-rich stars from \citet{Fernandez-Trincado2017a}, and 
	(\textit{iv}) Fifty (50) "bulge'' N-rich stars from \citet{Schiavon2017}.
	They are listed in Table \ref{Table1}, with their source catalog identified respectively as FT$+$16, M$+$16, FT$+$17 and S$+1$7,
	with the distances, proper motions and radial velocities.

\subsubsection{Complementary spectro-photometric distance estimates}

 \citet{Leung2019} derived precise distances\footnote{The astroNN catalog of distances for APOGEE DR16 stars is available at \url{https://www.sdss.org/dr16/data_access/value-added-catalogs/?vac_id=the-astronn-catalog-of-abundances,-distances,-and-ages-for-apogee-dr16-stars}} for the whole APOGEE DR16 spectra sample from a neural-network (deep learning) by training on the APOGEE-DR16/Gaia overlap. These distances have no Galaxy priors and are thus not biased by such. Figure \ref{Figure1} shows the comparison between \texttt{StarHorse} and \citet{Leung2019} distances (left panel), as well as the distance uncertainty as a function of the inferred distances (right panel). It is clear, that \texttt{StarHorse} distances have smaller uncertainties at larger ranges. Therefore we decide to use the \texttt{StarHorse} distances derived by \citet{Queiroz2019} as the main distance set for this work, also based on their better extinction treatment.

\begin{figure*}
	\begin{center}
		\includegraphics[width=190mm]{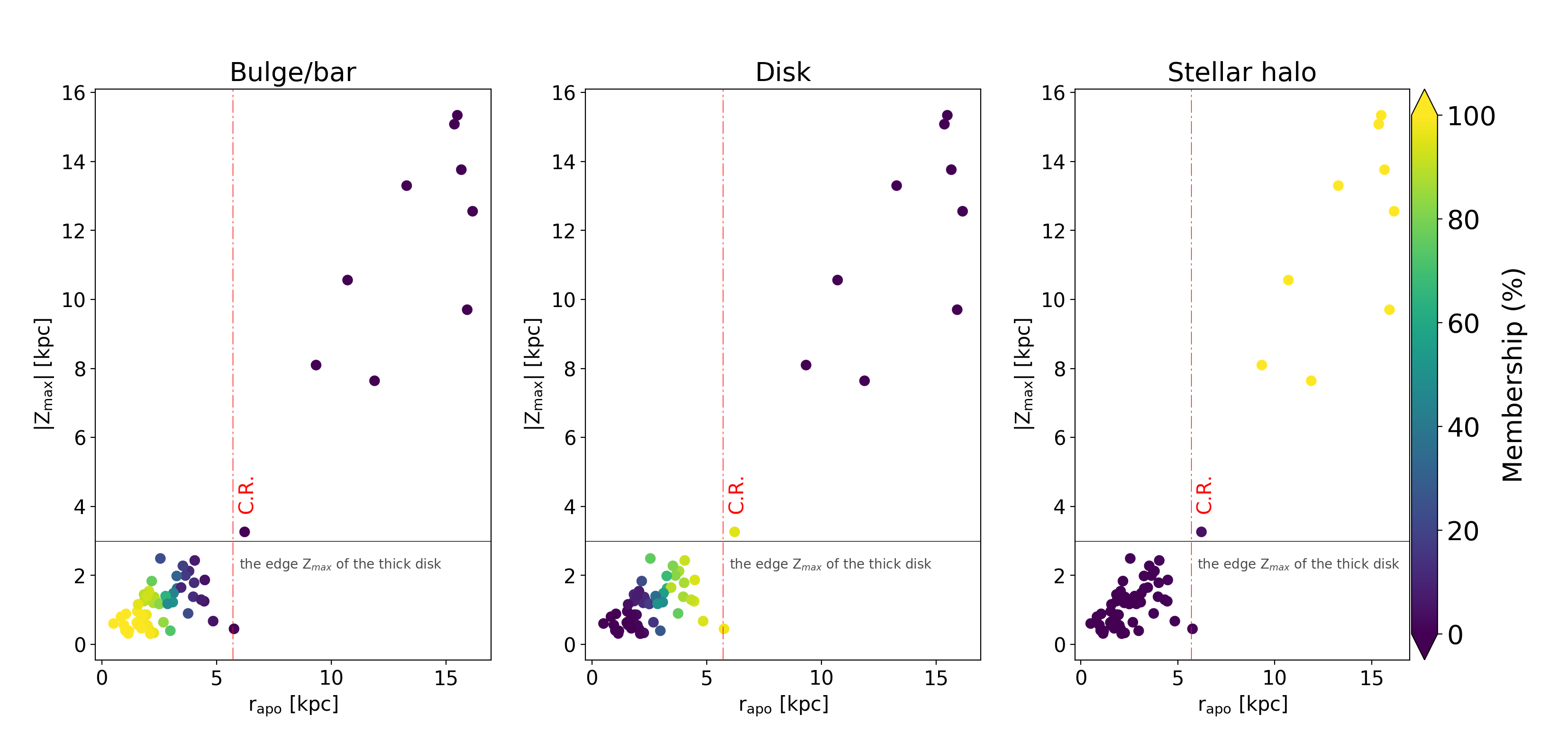}
		\caption{Membership probability adopting a bar pattern speed of 43 km s$^{-1}$ kpc$^{-1}$. The red dashdot line indicate the location of the bar's corotation radius (C.R. $=$ 5.7 kpc), whilst the black line represent the assumed edge Z$_{\rm max}$ of the thick disk \citep[$\sim$3 kpc, ;][]{Carollo2010}. From left to right panel shows the probability that each N-rich stars has to belong to the Bulge/bar, disk, and Stellar halo, respectively.}
		\label{Figure2}
	\end{center}
\end{figure*}

\section{The Milky Way model}
\label{section3}

In order to construct a comprehensive orbital study of selected N-rich stars across the Milky Way, we use a state-of-the art orbital integration model in a nonaxisymmetric gravitational potential, that fits the structural and dynamical parameters to the best we know of the recent knowledge of our Galaxy. 

For the computations in this work, we have employed the rotating "boxy/peanut" bar of the novel Galactic potential model called \texttt{GravPot16}\footnote{https://gravpot.utinam.cnrs.fr} along other composite stellar components. The considered structural parameters of our bar model, e.g., mass, present-day orientation and pattern speeds, is within observational estimations that lie in the range of 1.1$\times$10$^{10}$ M$_{\odot}$, 20$^{\circ}$ \citep[value adopted from dynamical constraints, as highlighted in Figure 12 in][]{Tang2018}, and 33--53 km s$^{-1}$ kpc in increments of 10 km s$^{-1}$ kpc$^{-1}$, respectively. The bar scale lengths are $x_0=$1.46 kpc, $y_{0}=$ 0.49 kpc, $z_0=$0.39 kpc and the middle region ends at the effective major semiaxis of the bar Rc $= 3.28$ kpc \citep{Robin2012}. The density-profile of the adopted "boxy/peanut" bar is exactly the same as in \citet{Robin2012}. In this paper, we explore a range of pattern speeds, which lead to different co-rotation radii for different pattern speeds, in particular, we compute the co-rotation radius, for the mass profiles of the static potential for any given pattern speed, $\Omega_{\rm bar}$,  such a patterns speed range places the bar's corotation radius in 4.6 kpc ($\Omega_{\rm bar} = 53$ km s$^{-1}$ kpc$^{-1}$), 5.7 kpc ( $\Omega_{\rm bar} = 43$ km s$^{-1}$ kpc$^{-1}$), and 7.2 kpc ($\Omega_{\rm bar} = 33 $ km s$^{-1}$ kpc$^{-1}$).

\texttt{GravPot16} considers on a global scale a 3D steady-state gravitational potential for the Milky Way, modelled as the superposition of axisymmetric and non-axysimmetric components. The axisymmetric potential is made-up of the superposition of many composite stellar populations belonging to seven thin discs, for each \textit{i}$^{th}$ component of the thin disc, we implemented an Einasto density-profile law \citet[][]{Einasto1979} as describe in \citet[][]{Robin2003}, superposed along with two thick disc components, each one following a simple hyperbolic secant squared decreasing vertically from the Galactic plane plus an exponential profile decreasing with Galactocentric radius as described in \citet{Robin2014}, we also implemented the density-profile of the interstellar matter (ISM) component with a density mass as presented in \citet{Robin2003}. 

The model, also correctly accounts for the underlying stellar halo, modelled by a Hernquist profile as already described in \citet{Robin2014}, and surrounded by a single spherical Dark Matter halo component \citet{Robin2003}, no time dependence of the density profiles is assumed. \texttt{GravPot16}  has been adopted in a score of papers \citep[e.g.,][]{ Recio-Blanco2017, Albareti2017, Helmi2018, Libralato2018, Schiappacasse-Ulloa2018, Tang2018, Tang2019, Minniti2018}. The mathematical formalism to derive the gravitational potential of the above Galactic component of the \texttt{GravPot16} algorithm, will be explained in full detail in a forthcoming paper (Fernandez-Trincado et al., in preparation).

For reference, the Galactic convention adopted by this work is: $X-$axis is oriented toward $l=$ 0$^{\circ}$ and $b=$ 0$^{\circ}$, and the $Y-$axis is oriented toward $l$ = 90$^{\circ}$ and $b=$0$^{\circ}$, and the disc rotates toward $l=$ 90$^{\circ}$; the velocity are also oriented in these directions. In this convention, the Sun's orbital velocity vector are [U$_{\odot}$,V$_{\odot}$,W$_{\odot}$] = [$11.1$, $12.24$, 7.25] km s$^{-1}$ \citep{Brunthaler2011}. The model has been rescaled to the Sun's galactocentric distance, 8.3 kpc, and the local rotation velocity of $239$ km s$^{-1}$.

For the computation of Galactic orbits of our N-rich stars, we have employed a simple Monte Carlo approach and the Runge-Kutta algorithm of seventh-eight order elaborated by \citet{fehlberg68}. The uncertainties in the input data (e.g., $\alpha$, $\delta$, distance, proper motions and line-of-sight velocity errors), were randomly propagated as 1$\sigma$ variation in a Gaussian Monte Carlo re-sampling. For each N-rich star we computed thousand orbits, computed backward in time during 3 Gyr. The average value of the orbital elements was found for thousand realizations, with uncertainty ranges given by the 16$^{\rm th}$ and 84$^{\rm th}$ percentile values.

In Table \ref{Table2}, we present the orbital parameters of our sample of N-rich stars for $\Omega_{\rm bar} = 43$ km s$^{-1}$ kpc$^{-1}$. We give the average value of the amplitude of the vertical oscillation ($|Z_{max}|$), the perigalactic distance, apogalactic distance, the eccentricity, the orbital Jacobi constant ($E_j$) computed in the reference frame of the bar, the 'characteristic' orbital energy as envisioned by \citep{Moreno2015}, the minimum and maximum of the \textit{z}-component of the angular momentum in the inertial frame, and the orbital behaviour. The errors provided in each column are computed as $\Delta = 0.5 \times $ (84$^{\rm th}$ percentile $-$ 16$^{\rm th}$ percentile). The number reported inside the parentheses indicate the standard deviation of the orbital elements when considering the three adopted bar pattern speeds, $\Omega_{\rm bar} = 33, 43, $ and $53$ km s$^{-1}$ kpc$^{-1}$. Most of the N-rich stars in our sample are giant stars located at large heliocentric distance $\gsim$ 5 kpc, and are found to have a radial orbit, which live in the inner Galaxy and the inner stellar halo (see Figure \ref{Figure2}) .   
 
\subsection{Limitations of our model}

The more important limitations of our model are: 

\begin{itemize}
	\item[I.] We ignore secular changes in the Milky Way potential over time, which are expected although our Galaxy has had a quiet recent accretion history
	\item[II.] We do not consider  perturbations due to spiral arms, an in-depth analysis is beyond the scope of this paper.  
\end{itemize}

\section{Orbital elements}
\label{section4}

	In this section we briefly describe the main orbital elements of the 64 N-rich stars in our sample, which are listed in Table \ref{Table2}. In basis to these orbital elements we calculate the membership probability that each star has to belong to each Galactic component, as will be discussed in the next section. It is important to note that we have calculate the minimum and maximum value of the z-component of the angular momentum ($L_{z}$) in the inertial frame along the whole integration time, as this quantity is not conserved in a model with non-axisymmetric structures. We are interested only in the sign, in order to verify whether the orbital motion of the N-rich stars has a prograde or a retrograde sense with respect to the Galactic rotation. Regarding, the different values of $\Omega_{\rm bar}$, we can see that most of orbits are not sensitive to the change of this parameter.
	
	We find that the vast majority ( $\sim83\%$ ) of the N-rich stars have a radial orbit, with an apocentre $\lesssim 5.7$ kpc (inside the corotation radius, C.R.), and $Z_{\rm max}\lesssim 3$ kpc, placing them at the edge of the inner disk and/or bulge/bar region, therefore one would expect that most of the N-rich stars follow the typical orbital configuration of the bulge/bar structure. Furthermore, the orbital elements, presented in Table \ref{Table2}, show that 53\% of the N-rich stars exhibit orbits that change their sense of motion from prograde to retrograde (and vice verse), and are confined or belong to the inner Galaxy. This effect is produced by the presence of the bar, likely indicative of an early dynamics phase of the evolution of the central region of the Galaxy, meaning that such objects were likely formed in a very early stage of the Galaxy \citep[see, e.g.,][]{Bekki2019}, probably before bar formation, and a few of them were trapped by the bar structure later on. As for the rest of the stars (39\%) in our sample exhibit prograde orbits with respect to the direction of the Galactic rotation and lives in the inner Galaxy, they are very likely members of the bulge/bar structure, as seen in Figure \ref{Figure2} (left panel), whilst a few N-rich stars (8 \%) have a retrograde sense with respect to the Galactic rotation and lies in orbits with larger apogalactocentric and $Z_{\rm max}$ distances, they are likely members of the stellar halo component as seen in Figure \ref{Figure2} (right panel).
	
	In Table \ref{Table2}, we can see that most of the stellar orbits (95\%) become highly eccentric ($e$ $\gsim$ 0.5). In other words, such stars appear to behave as bulge/bar-like and halo-like orbits. There are nonetheless, ten stars with a different behaviour, as seen in Figure \ref{Figure2}, on highly elliptical orbits with eccentricities of $e \gsim 0.5$, reaching out to a maximum distance from the Galactic plane $|Z_{max}|$ larger than 3 kpc, and orbital oscillations beyond C.R. Such stars appear to behave as halo-like stars. 
	
     Now, we turn our attention to 2M17180311$-$2750124, 2M18372953$-$2911046 and 2M17504980$-$2255083, which show in-plane prograde trajectories ($|Z_{max}| \lsim$ 1.17 and 1.64 kpc, respectively) and lower eccentricities ($\lsim 0.5$). Among they, we have identified to 2M18372953$-$2911046 to be in the boundary between two Galactic components (bulge/bar and disk structure), whilst 2M17180311$-$2750124 and 2M17504980$-$2255083 lives in the inner Galaxy, and have high probability to support the inner disk structure inside C.R. and appear to behave as a disk-like orbit. We find that the orbit of 2M17180311$-$2750124 has energies allowing they to move around C.R. radius, this is the star in $r_{\rm apo}\sim$ 5.8 kpc and Z$_{\rm max}\sim$0.45 in Figure \ref{Figure2}.
	
	Among the metal-poor N-rich stars is TYC 5619-109-1, an exotic 'early-AGB' star \citep[e.g.,][]{Pereira2017} identified by \citet[][]{Fernandez-Trincado2016b}, which is the only known N-rich star with extreme enrichment of light-/heavy-elements. TYC 5619-109-1 lies in a retrograde and eccentric orbit (\text{e}$>$0.59) with high vertical excursions ($|Z_{max}| \lsim 7.6$ kpc) from the plane, indicating that it belong to the halo-like population. Figure \ref{Figure2} (right panel) shows that TYC 5619-109-1 have high probability to belong to the inner halo. Thus, the extreme [Al/Fe] and [N/Fe] abundances \citep[e.g.,][]{Fernandez-Trincado2016b, Pereira2017} well above  the typical Galactic levels and the retrograde and halo nature of its orbit, may be interpreted as possibly being tidal debris candidate of the unusual globular cluster $\omega$ Cen.
	
	Finally, we identify one star as a halo interloper in the Galactic bulge,  (2M18550318$-$3043368) from \citet{Schiavon2017}, which has a large apogalacentric distance $\gsim 10.7$ kpc and a high vertical excursion from the Galactic plane ($\lsim10.6$ kpc), currently passing through the bulge/bar component of the Milky Way, and as we will see soon, this star have a high probability to belong to the stellar halo (see Table \ref{Table3}). It is likely that this star is a halo interloper that happens to be at same distance and location than bulge N-rich stars, suggesting that contamination from (inner) halo stars is relevant when attempting to trace this anomalous population toward the inner Galaxy. However, there is not any obvious dependence among the metallicity and orbital properties for such objects, therefore it is probable that they were formed during the very early stages of the evolution of the Milky Way.

\begin{figure}
	\begin{center}
		\includegraphics[width=90mm]{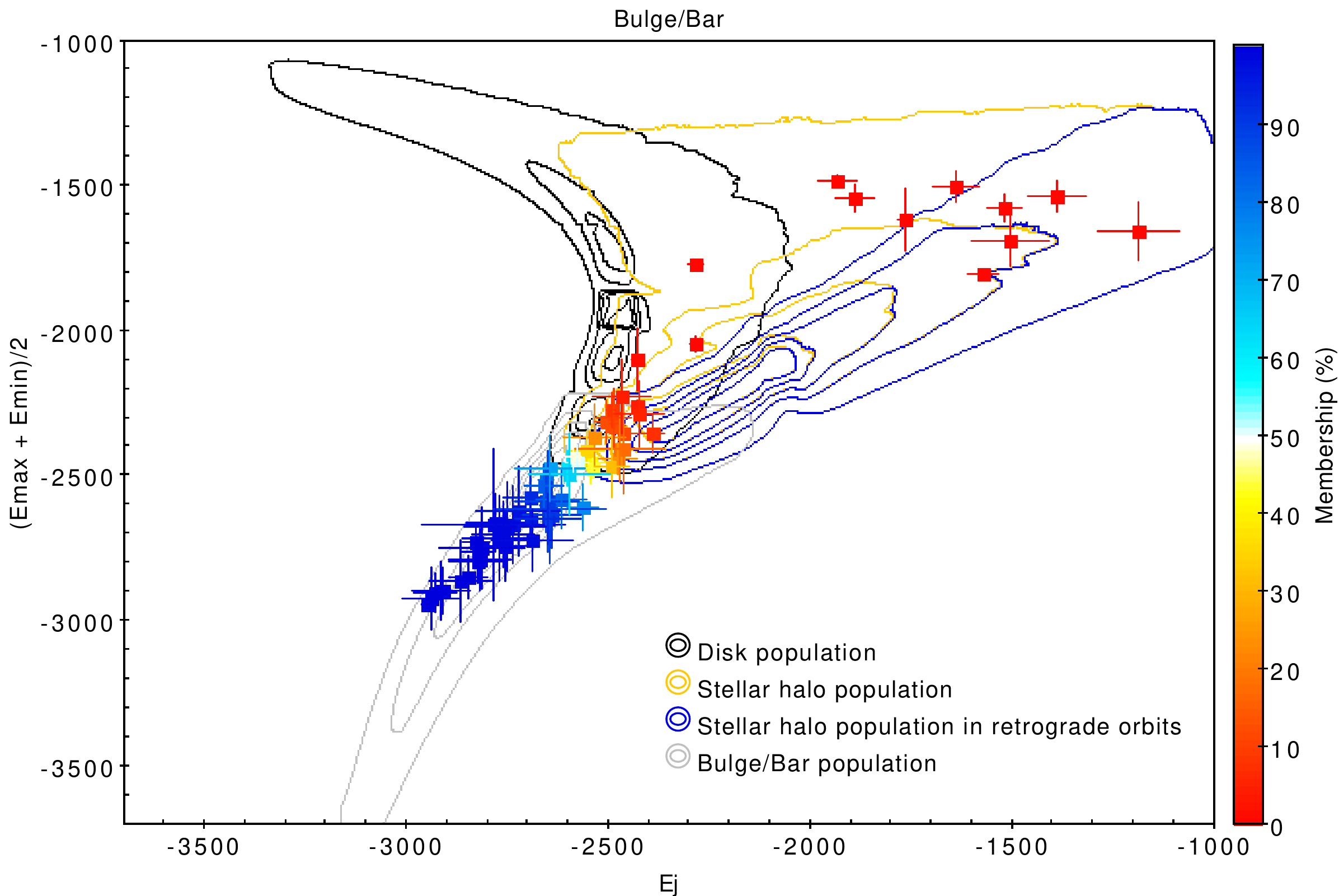}
		\includegraphics[width=90mm]{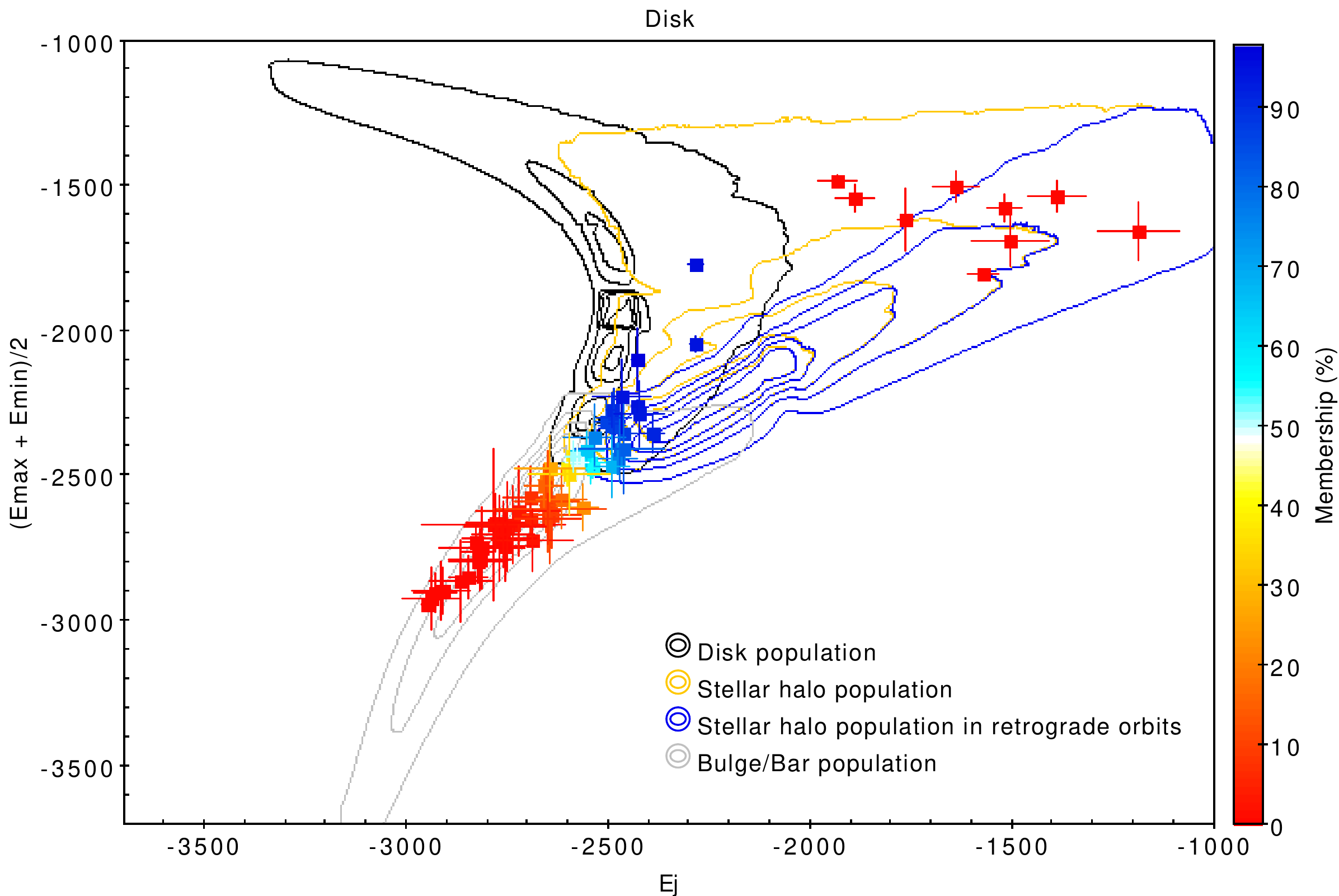}
		\includegraphics[width=90mm]{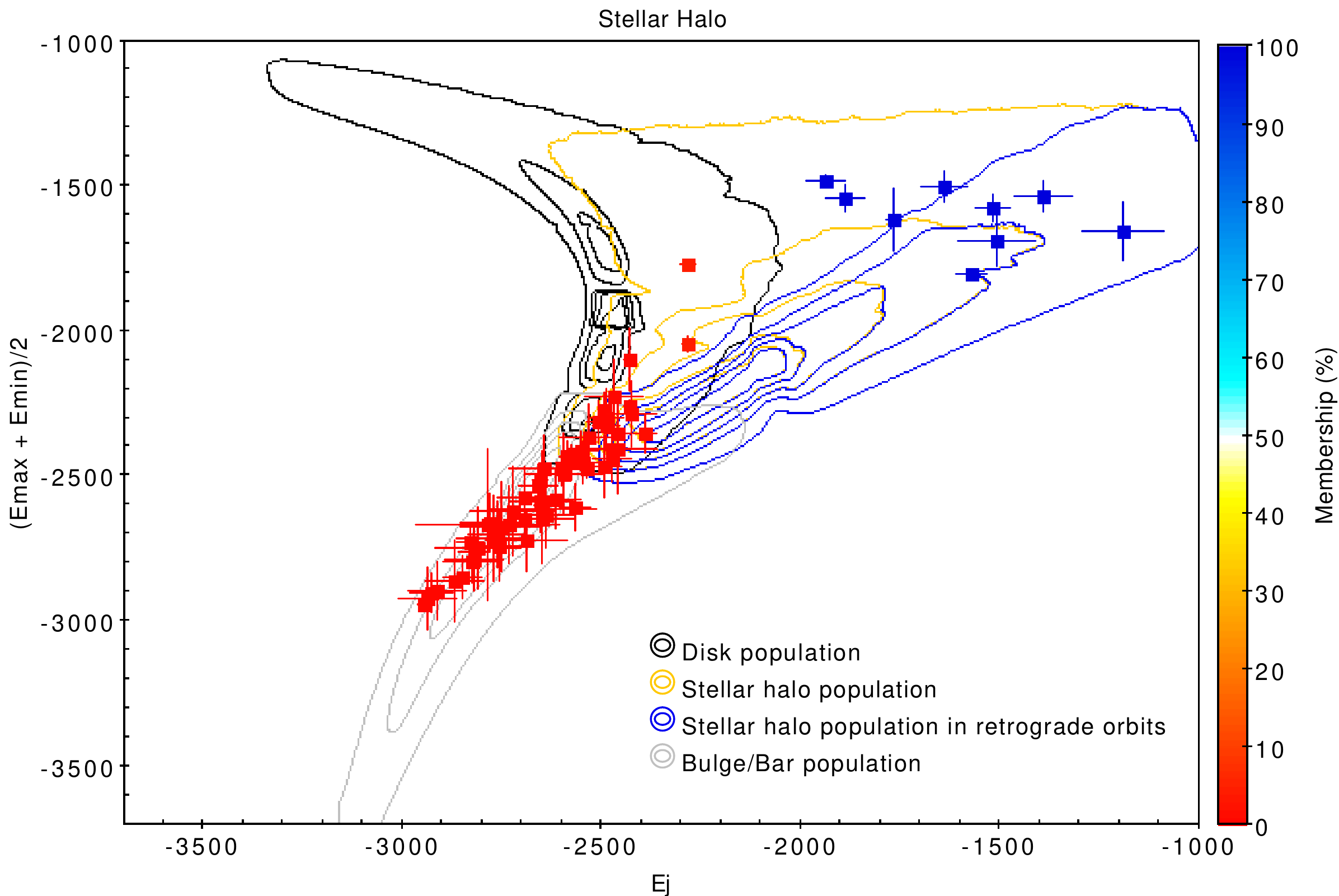}
		\caption{The 'characteristic' orbital energy, E$_{\rm char}$, versus the orbital Jacobi constant ($E_{j}$), in units of 10$^{2}$ km$^{2}$ s$^{2}$, computed in the reference frame of the bar for $\Omega_{\rm bar} = 43 $ km s$^{-1}$ kpc$^{-1}$. From top to bottom panel show the probability that each N-rich star has to belong to the bulge/bar (top), disk (middle), and stellar halo (bottom), respectively. Isophotes represent the synthetic stellar populations (bulge/bar: grey; disk: black; canonical stellar halo: orange and stellar halo in retrograde sense) from a set of high resolution simulations from \texttt{GravPot16} with varying bar pattern speeds (Fern\'andez-Trincado et al. in preparation).}
		\label{Figure3}
	\end{center}
\end{figure}

\section{Dynamical orbital classification}
\label{section5}

Following a similar approach to the one recently presented in \citet{Perez-Villegas2020} to classify globular cluster orbits, we employ a more robust method to quantitatively separate the N-rich stars into different stellar populations. To do this, we employ a set of test particles evolved in a nonaxisymmetric galactic potential with varying bar pattern speeds (33, 43, and 53 km s$^{-1}$ kpc$^{-1}$) in \texttt{GravPot16} and in statistical equilibrium with the potential imposed (Fern\'andez-Trincado et al. in preparation), which have inherited the information on both density and kinematics. Our simulations aims to mimic the present spatial and kinematic distribution of the bar/bulge, disk and stellar halo populations. The simulations were ran employing the same Galactic configuration as mentioned in \S\ref{section3}. The initial conditions follow the density profiles and kinematics from the updated version of the Besan\c{c}on galaxy model with parameters set in \citet{Robin2003, Robin2012, Robin2014, Robin2017}.
	
	For each synthetic stellar population (SSP), we employ the 'characteristic' orbital energy (${\rm E_{char}} = \frac{\rm E_{\rm max} + E_{\rm min}}{2}$) as envisioned by \citet{Moreno2015} and the orbital Jacobi constant (${\rm E_j}$), as illustrated in Figure \ref{Figure3} in different density contours ({\bf grey:} bulge/bar; {\bf black:} disk; {\bf orange and blue:} stellar halo population in prograde and retrograde orbits, respectively). We limit our analysis to SSP with $r_{apo} < 50$ kpc. For the SSP separation, we take into account the contribution of the set of orbits of each SSP and the three bar pattern speeds. From each SSP, we have applied a 2D Gaussian Mixture Method (GMM, Pedregosa et al. 2011) to obtain the centre $G_i({\rm E_{char}}, {\rm E_j})$, with $\sigma_i({\rm E_{char}}, {\rm E_j})$, and weight $P_i$. For the bulge/bar component $G_{bulge/bar}(-2697.87, -2754.07)$, $\sigma_{bulge/bar}(300.58, 143.49)$, and $P_{bulge/bar} = 0.26$; for the disk $G_{disk}(-1922.49, -2610.21)$, $\sigma_{disk}(306.57, 195.17)$, and $P_{disk} = 0.73$ and for the stellar halo $G_{halo}(-1972.84, -2071.58)$, $\sigma_{halo}(293.56, 373.94)$, and $P_{halo} = 0.01$. With the information of the each component provided by the SSP, we calculate the membership probability that each N-rich star has to belong to each Galactic component, the results are listed in Table \ref{Table3}. 
	
	With the information of each SSP and using a Gaussian distribution probability, we compute the membership probability each N-rich star to each Galactic component. The maximum value of probability is listed in Table \ref{Table3} and is used to classify the N-rich stars into each Galactic component. Figure \ref{Figure3} shows the membership probability for $\Omega_{\rm bar} = 43$ km s$^{-1}$ kpc$^{-1}$, the bluer/redder the dots the more/less probable the membership to the 
	bulge/bar (top panel), disk (middle panel), and stellar halo (bottom panel) is. Dark blue corresponds to a probability higher than 80\% and dark red to one lower than 10\%, whilst cyan, yellow and orange dots mark probabilities between 10\% to 80\%, i.e. stars in the boundary between two SSPs. Figure \ref{Figure3} also shows a clear correlation between ${\rm E_{char}}$ and ${\rm E_j}$, with the vast majority of the N-rich stars belonging to the bulge/bar component (where we can note that bulge/bar population and N-rich stars overlap in the innermost part of the Galaxy) and the stellar halo (many of which lie in retrograde orbits).  

\subsection{Membership probability}

In our sample, 50 out of 64 N-rich stars are identified as bulge stars in \citet{Schiavon2017}, and based on their dynamical properties, for the bar pattern speed of 43 km s$^{-1}$ kpc$^{-1}$, we found that 39 of them, have characteristics of bulge/bar population with a higher probability (membership probability $>80\%$), whereas the other 11 stars appear to be intruders from other Galactic components, that currently are crossing the inner parts of the Milky Way Galaxy, including a clear halo interloper (e.g., 2M18550318$-$3043368). 

Additionally, the 4/5 N-rich stars reported in \citep{Martell2016} and TYC 5619-109-1 \citep{Fernandez-Trincado2016b} are identified as part of the stellar halo in our classification with a higher probability (membership probability of 100$\%$). There are $\sim$17 stars, that change of component with the bar pattern speed, and there are N-rich stars with significant probability to be in two Galactic components (see Table \ref{Table3}), which are in the boundary between bulge/bar and disk and/or disk and (inner) stellar halo, one example is 2M16493657$-$2028146 from \citet{Schiavon2017}, which is a N-rich star in the boundary between bulge/bar and disk component (with probability between 10\% to 80\%).  

\section{Concluding remarks}
\label{section6}

In this work we exploited the accurate data from the European Space Agency's \texttt{Gaia} mission Second Data Release (DR2) and the APOGEE-2 survey to present, for the first time, a dynamical characterization of giant stars with anomalously high levels of [N/Fe] in the Milky Way. This analysis have been carried out with the novel Milky Way model called \texttt{GravPot16}, where the Galactic orbits have been integrated in a nonaxisymmetric configuration, including a rotating bar potential, where we vary the angular velocity of the bar. With the orbital properties of synthetic stellar populations and observed stars, a characteristic orbital energy and the orbital Jacobi constant have been computed. We applied a simple probabilistic approach to estimate the membership probability that each star has to be a member of different stellar populations: bulge/bar, disk and/or  (inner) stellar halo. 

We find that a significant fraction (53\%, 34 out of 64 stars) of the selected N-rich stars have orbits in a retrograde/prograde (vise verse) sense with respect to the direction of the Galactic rotation due to the presence of the bar structure, and currently lives in the inner Galaxy. Among this sub-sample, we find that 28 out of 34 of them have high probability ($>80$\%) to belong to the bulge/bar component, whilst 6 out of 34 are in the boundary between two Galactic components. 

We have identified five (8\%) of field N-rich stars in high eccentric ($>0.5$) and retrograde orbits. In termns of dynamical properties, the orbital characteristics of the these stars identified to be part of the inner halo ($r_{gal}<$ 20 kpc) with higher probability ($>98$\%) according to the present classification criteria. 

Among the N-rich stars, we also identified a considerable number (39\%, 25 out of 64 stars) of them in prograde orbits. From this sub-group, we identified 4 out of 25 stars with orbits in high eccentricities ($>$0.65) and with high probability ($>90$\%) to belong to the stellar halo, and 10 out of 25 have a high probability ($>80$\%) to belong to the inner disk component, whilst 11 out of 25 lives in the inner Galaxy, but have dynamical properties that are in the boundary between two Galactic components, i.e., the bulge/bar and/or disk population, depending on the bar angular velocity, $\Omega_{\rm bar}$.

Lastly, we confirm the halo nature of the unusual N-rich star, TYC 5619-109-1, which lie in a retrograde orbital configuration and with high probability to belong to the halo-like population. Both, chemistry and dynamical properties may favour the hypothesis of the ejection from a globular cluster, and it is likely that TYC 5619-109-1 share similar genetic similarities with the globular cluster $\omega$ Cen. 


\section*{Acknowledgements}

We acknowledge the anonymous referee for enlightening comments that greatly improved this paper. 

 J.G.F-T is supported by FONDECYT No. 3180210 and Becas Iberoam\'erica Investigador 2019, Banco Santander Chile. J.G.F-T is grateful to Friedrich Anders for his precious help with \texttt{StarHorse}.
 
 E.M acknowledge support from UNAM/PAPIIT grant IN105916. 
 
 A.P-V acknowledges a FAPESP for the postdoctoral fellowship grant no. 2017/15893-1 and the DGAPA-PAPIIT grant IG100319. 

  L.C-V thanks the Fondo Nacional de Financiamiento para la Ciencia, La Tecnolog\'ia y la innovaci\'on "FRANCISCO JOS\'E DE CALDAS", MINCIENCIAS, and the VIIS for the economic support of this research.

   Funding for the \texttt{GravPot16} software has been provided by the Centre national d'\'etudes spatiales (CNES) through grant 0101973 and UTINAM Institute of the Universit\'e de Franche-Comt\'e, supported by the R\'egion de Franche-Comt\'e and Institut des Sciences de l'Univers (INSU). Simulations have been executed on computers from the Utinam Institute of the Universit\'e de Franche-Comt\'e, supported by the R\'egion de Franche-Comt\'e and Institut des Sciences de l'Univers (INSU), and on the supercomputer facilities of the M\'esocentre de calcul de Franche-Comt\'e. 

This work presents results from the European Space Agency (ESA) space mission Gaia. Gaia data are being processed by the Gaia Data Processing and Analysis Consortium (DPAC). Funding for the DPAC is provided by national institutions, in particular the institutions participating in the Gaia MultiLateral Agreement (MLA). The Gaia mission website is \url{https://www.cosmos.esa.int/gaia}. The Gaia archive website is \url{https://archives.esac.esa.int/gaia}.

Funding for the Sloan Digital Sky Survey IV has been provided by the Alfred P. Sloan Foundation, the U.S. Department of Energy Office of Science, and the Participating Institutions. SDSS- IV acknowledges support and resources from the Center for High-Performance Computing at the University of Utah. The SDSS web site is www.sdss.org. SDSS-IV is managed by the Astrophysical Research Consortium for the Participating Institutions of the SDSS Collaboration including the Brazilian Participation Group, the Carnegie Institution for Science, Carnegie Mellon University, the Chilean Participation Group, the French Participation Group, Harvard-Smithsonian Center for Astrophysics, Instituto de Astrof\`{i}sica de Canarias, The Johns Hopkins University, Kavli Institute for the Physics and Mathematics of the Universe (IPMU) / University of Tokyo, Lawrence Berkeley National Laboratory, Leibniz Institut f\"{u}r Astrophysik Potsdam (AIP), Max-Planck-Institut f\"{u}r Astronomie (MPIA Heidelberg), Max-Planck-Institut f\"{u}r Astrophysik (MPA Garching), Max-Planck-Institut f\"{u}r Extraterrestrische Physik (MPE), National Astronomical Observatory of China, New Mexico State University, New York University, University of  Dame, Observat\'{o}rio Nacional / MCTI, The Ohio State University, Pennsylvania State University, Shanghai Astronomical Observatory, United Kingdom Participation Group, Universidad Nacional Aut\'{o}noma de M\'{e}xico, University of Arizona, University of Colorado Boulder, University of Oxford, University of Portsmouth, University of Utah, University of Virginia, University of Washington, University of Wisconsin, Vanderbilt University, and Yale University.





\bibliographystyle{mnras}
\bibliography{references} 





\twocolumn

\clearpage
\onecolumn
\small\addtolength{\tabcolsep}{-0.5pt}
\begin{longtable}{| l | c | c | c | c | c |}
\caption{N-rich stars data.}
\label{Table1}

\\
\hline
   APOGEE\_ID & $d_{\odot}$ & $V_r$ & $\mu_{\alpha} \times \cos{\delta}$ & $\mu_{\delta}$  & Source\\
    &  (kpc) & $(\mathrm{km}\,\mathrm{s}^{-1})$ & $(\mathrm{mas}\,\mathrm{yr}^{-1})$ & $(\mathrm{mas}\,\mathrm{yr}^{-1})$ & \\ 
    \endfirsthead
\hline
\hline
    2M17535944$+$4708092 &      13.92 $\pm$ 1.26 &    -265.6 $\pm$ 0.1  &     -1.2  $\pm$ 0.05  &     -2.07  $\pm$ 0.05 &  FT$+$17 \\
    2M17350460$-$2856477 &      4.97  $\pm$ 0.59 &    -106.5 $\pm$ 0.39 &     -0.41 $\pm$ 0.31  &     -9.89  $\pm$ 0.24 &  FT$+$17 \\
    2M12155306$+$1431114 &      13.83 $\pm$ 1.26 &    100.1  $\pm$ 0.12 &     -1.01 $\pm$ 0.12  &     -1.42  $\pm$ 0.11 &  FT$+$17 \\
    2M16062302$-$1126161 &      3.2   $\pm$ 0.34 &    -105.9 $\pm$ 0.33 &     -6.37 $\pm$ 0.1   &     -9.04  $\pm$ 0.06 &  FT$+$17 \\
    2M17454705$-$2639109 &      6.15  $\pm$ 1.57 &    -74.9  $\pm$ 0.01 &     -0.53 $\pm$ 0.26  &     -5.53  $\pm$ 0.23 &  FT$+$17 \\
    2M17492967$-$2328298 &      4.71  $\pm$ 0.67 &    26.3   $\pm$ 0.04 &     0.89  $\pm$ 0.15  &     -8.78  $\pm$ 0.11 &  FT$+$17 \\
    2M17534571$-$2949362 &      5.38  $\pm$ 0.86 &    -140.4 $\pm$ 0.0  &     -4.89 $\pm$ 0.14  &     -4.76  $\pm$ 0.12 &  FT$+$17 \\
    2M17180311$-$2750124 &      3.35  $\pm$ 0.62 &    -113.8 $\pm$ 0.1  &     -3.13 $\pm$ 0.08  &     -5.46  $\pm$ 0.06 &  FT$+$17 \\
    2M02491285$+$5534213 &      1.26  $\pm$ 0.07 &    -223.2 $\pm$ 0.14 &     41.09 $\pm$ 0.08  &     0.1    $\pm$ 0.08 &  FT$+$17 \\
    2M15113526$+$3551140 &      13.96 $\pm$ 1.29 &    -246.4 $\pm$ 0.06 &     -2.14 $\pm$ 0.05  &     -1.83  $\pm$ 0.05 &  M$+$16  \\
    2M15204588$+$0055032 &      17.86 $\pm$ 1.95 &    -56.2  $\pm$ 0.06 &     -2.79 $\pm$ 0.07  &     -2.69  $\pm$ 0.06 &  M$+$16  \\
    2M13251355$-$0044438 &      13.73 $\pm$ 1.52 &    -99.3  $\pm$ 0.16 &     -3.47 $\pm$ 0.08  &     -3.59  $\pm$ 0.06 &  M$+$16  \\
    2M17252263$+$4903137 &      16.13 $\pm$ 1.69 &    -249.5 $\pm$ 0.65 &     -1.56 $\pm$ 0.05  &     -0.69  $\pm$ 0.05 &  M$+$16  \\
    2M16011638$-$1201525 &      5.72  $\pm$ 0.31 &    83.9   $\pm$ 0.6  &     -11.7 $\pm$ 0.1   &     -18.0  $\pm$ 0.07 &  FT$+$16 \\
    2M16493657$-$2028146 &      7.95  $\pm$ 0.89 &    65.1   $\pm$ 1.02 &     -5.38 $\pm$ 0.06  &     -3.02  $\pm$ 0.05 &  S$+$17  \\
    2M16514646$-$2127071 &      4.79  $\pm$ 0.48 &    54.7   $\pm$ 0.0  &     -5.69 $\pm$ 0.06  &     -6.06  $\pm$ 0.05 &  S$+$17  \\
    2M17024730$-$2210387 &      5.21  $\pm$ 0.67 &    -21.1  $\pm$ 0.01 &     -5.28 $\pm$ 0.08  &     -2.67  $\pm$ 0.05 &  S$+$17  \\
    2M17134700$-$2441353 &      7.49  $\pm$ 0.92 &    -61.7  $\pm$ 0.02 &     -8.13 $\pm$ 0.1   &     -6.58  $\pm$ 0.07 &  S$+$17  \\
    2M17161691$-$2458586 &      6.45  $\pm$ 1.04 &    93.3   $\pm$ 0.0  &     -1.79 $\pm$ 0.09  &     -2.42  $\pm$ 0.07 &  S$+$17  \\
    2M17173203$-$2439094 &      6.71  $\pm$ 0.85 &    5.6    $\pm$ 0.0  &     -4.08 $\pm$ 0.15  &     -9.02  $\pm$ 0.1  &  S$+$17  \\
    2M17193271$-$2732214 &      7.68  $\pm$ 0.78 &    183.0  $\pm$ 0.0  &     -1.76 $\pm$ 0.1   &     -4.94  $\pm$ 0.08 &  S$+$17  \\
    2M17205201$-$2903061 &      6.58  $\pm$ 0.75 &    -7.6   $\pm$ 0.45 &     -2.22 $\pm$ 0.14  &     -6.01  $\pm$ 0.1  &  S$+$17  \\
    2M17211817$-$2735530 &      6.94  $\pm$ 0.49 &    -10.0  $\pm$ 0.0  &     -3.03 $\pm$ 0.18  &     -7.7   $\pm$ 0.13 &  S$+$17  \\
    2M17263951$-$2406247 &      7.0   $\pm$ 0.78 &    -51.3  $\pm$ 0.01 &     -3.64 $\pm$ 0.17  &     -5.5   $\pm$ 0.12 &  S$+$17  \\
    2M17271907$-$2718040 &      4.79  $\pm$ 0.56 &    63.8   $\pm$ 0.03 &     -9.02 $\pm$ 0.22  &     -7.03  $\pm$ 0.16 &  S$+$17  \\
    2M17303980$-$2330234 &      7.33  $\pm$ 0.83 &    -15.0  $\pm$ 0.0  &     -5.52 $\pm$ 0.12  &     -1.1   $\pm$ 0.09 &  S$+$17  \\
    2M17305251$-$2651528 &      6.77  $\pm$ 0.72 &    42.9   $\pm$ 0.0  &     1.07  $\pm$ 0.22  &     -7.97  $\pm$ 0.17 &  S$+$17  \\
    2M17334208$-$2958347 &      7.44  $\pm$ 0.73 &    90.4   $\pm$ 0.71 &     -2.48 $\pm$ 0.35  &     -4.27  $\pm$ 0.28 &  S$+$17  \\
    2M17343610$-$2909472 &      7.07  $\pm$ 0.74 &    40.1   $\pm$ 0.31 &     -5.73 $\pm$ 0.57  &     -3.89  $\pm$ 0.46 &  S$+$17  \\
    2M17343654$-$1956596 &      5.84  $\pm$ 0.75 &    -6.8   $\pm$ 0.08 &     -6.49 $\pm$ 0.1   &     -1.41  $\pm$ 0.08 &  S$+$17  \\
    2M17343807$-$2557555 &      6.64  $\pm$ 0.58 &    -84.5  $\pm$ 0.01 &     -4.19 $\pm$ 0.22  &     -6.22  $\pm$ 0.18 &  S$+$17  \\
    2M17352288$-$2913255 &      5.99  $\pm$ 0.97 &    47.0   $\pm$ 0.81 &     -3.45 $\pm$ 0.34  &     -5.19  $\pm$ 0.25 &  S$+$17  \\
    2M17353215$-$2759106 &      7.29  $\pm$ 0.73 &    18.4   $\pm$ 0.17 &     -1.05 $\pm$ 0.37  &     -3.51  $\pm$ 0.31 &  S$+$17  \\
    2M17354267$-$2406233 &      7.59  $\pm$ 0.16 &    -21.3  $\pm$ 0.0  &     -2.11 $\pm$ 0.34  &     1.16   $\pm$ 0.21 &  S$+$17  \\
    2M17382269$-$2748001 &      7.39  $\pm$ 0.71 &    -123.4 $\pm$ 0.21 &     -4.07 $\pm$ 0.21  &     -7.2   $\pm$ 0.15 &  S$+$17  \\
    2M17390422$-$2943520 &      7.32  $\pm$ 1.19 &    52.9   $\pm$ 0.06 &     0.25  $\pm$ 0.54  &     -9.79  $\pm$ 0.4  &  S$+$17  \\
    2M17404143$-$2714570 &      6.7   $\pm$ 0.68 &    -74.3  $\pm$ 0.03 &     -6.26 $\pm$ 0.37  &     -8.08  $\pm$ 0.28 &  S$+$17  \\
    2M17434675$-$2616068 &      6.71  $\pm$ 0.75 &    175.7  $\pm$ 0.01 &     -9.29 $\pm$ 0.2   &     -5.56  $\pm$ 0.15 &  S$+$17  \\
    2M17442343$-$2627304 &      8.12  $\pm$ 0.89 &    -209.6 $\pm$ 0.0  &     -4.17 $\pm$ 0.22  &     -6.13  $\pm$ 0.16 &  S$+$17  \\
    2M17453131$-$2342147 &      7.78  $\pm$ 0.82 &    97.9   $\pm$ 0.1  &     0.12  $\pm$ 0.14  &     -4.41  $\pm$ 0.1  &  S$+$17  \\
    2M17464449$-$2531533 &      7.9   $\pm$ 0.83 &    -36.6  $\pm$ 0.07 &     -5.41 $\pm$ 0.28  &     -7.47  $\pm$ 0.23 &  S$+$17  \\
    2M17480576$-$2445000 &      6.47  $\pm$ 0.66 &    -76.7  $\pm$ 0.07 &     -0.44 $\pm$ 0.41  &     -3.51  $\pm$ 0.32 &  S$+$17  \\
    2M17482995$-$2305299 &      7.02  $\pm$ 0.56 &    -216.5 $\pm$ 0.0  &     -1.07 $\pm$ 0.11  &     -6.56  $\pm$ 0.09 &  S$+$17  \\
    2M17494963$-$2318560 &      7.16  $\pm$ 0.66 &    -42.5  $\pm$ 0.0  &     0.04  $\pm$ 0.15  &     -4.22  $\pm$ 0.12 &  S$+$17  \\
    2M17504980$-$2255083 &      5.0   $\pm$ 0.64 &    47.1   $\pm$ 0.0  &     -6.62 $\pm$ 0.12  &     0.32   $\pm$ 0.1  &  S$+$17  \\
    2M17523300$-$3027521 &      8.27  $\pm$ 1.09 &    152.9  $\pm$ 0.28 &     -4.39 $\pm$ 0.23  &     -12.31 $\pm$ 0.19 &  S$+$17  \\
    2M17554454$-$2123058 &      4.78  $\pm$ 0.79 &    94.8   $\pm$ 0.02 &     -6.15 $\pm$ 0.22  &     -8.87  $\pm$ 0.2  &  S$+$17  \\
    2M18014817$-$3026237 &      8.15  $\pm$ 0.79 &    -20.2  $\pm$ 0.3  &     -4.34 $\pm$ 0.09  &     -6.73  $\pm$ 0.08 &  S$+$17  \\
    2M18022530$-$2928338 &      5.14  $\pm$ 0.7  &    157.5  $\pm$ 0.0  &     -3.27 $\pm$ 0.23  &     -3.12  $\pm$ 0.18 &  S$+$17  \\
    2M18032356$-$3001588 &      5.97  $\pm$ 0.54 &    -13.8  $\pm$ 2.14 &     2.51  $\pm$ 0.1   &     -6.51  $\pm$ 0.08 &  S$+$17  \\
    2M18033335$-$2929122 &      7.72  $\pm$ 0.8  &    -55.9  $\pm$ 0.37 &     -1.13 $\pm$ 0.13  &     -4.46  $\pm$ 0.1  &  S$+$17  \\
    2M18035944$-$2908195 &      6.85  $\pm$ 0.1  &    172.7  $\pm$ 0.0  &     -7.93 $\pm$ 0.1   &     -4.37  $\pm$ 0.09 &  S$+$17  \\
    2M18054875$-$3122407 &      8.03  $\pm$ 0.79 &    3.4    $\pm$ 0.48 &     -0.54 $\pm$ 0.1   &     -5.08  $\pm$ 0.08 &  S$+$17  \\
    2M18061336$-$3147053 &      6.61  $\pm$ 0.89 &    -57.5  $\pm$ 0.32 &     -1.05 $\pm$ 0.08  &     -7.26  $\pm$ 0.07 &  S$+$17  \\
    2M18090957$-$1559276 &      8.05  $\pm$ 1.35 &    -8.7   $\pm$ 0.0  &     -1.24 $\pm$ 0.39  &     -3.38  $\pm$ 0.37 &  S$+$17  \\
    2M18102953$-$2707208 &      6.42  $\pm$ 0.7  &    -51.9  $\pm$ 0.0  &     -7.51 $\pm$ 0.12  &     -7.33  $\pm$ 0.1  &  S$+$17  \\
    2M18120031$-$1350169 &      6.06  $\pm$ 0.63 &    -36.2  $\pm$ 0.01 &     -1.49 $\pm$ 0.27  &     -7.73  $\pm$ 0.24 &  S$+$17  \\
    2M18121957$-$2926310 &      8.11  $\pm$ 1.02 &    45.2   $\pm$ 0.01 &     -1.83 $\pm$ 0.07  &     -4.32  $\pm$ 0.07 &  S$+$17  \\
    2M18124455$-$2719146 &      6.68  $\pm$ 0.84 &    -156.9 $\pm$ 0.0  &     -4.47 $\pm$ 0.13  &     -4.01  $\pm$ 0.11 &  S$+$17  \\
    2M18165340$-$2017051 &      5.99  $\pm$ 0.91 &    70.0   $\pm$ 0.09 &     -4.58 $\pm$ 0.32  &     -5.94  $\pm$ 0.28 &  S$+$17  \\
    2M18334592$-$2903253 &      5.6   $\pm$ 0.55 &    -152.2 $\pm$ 0.01 &     -3.31 $\pm$ 0.06  &     -5.79  $\pm$ 0.05 &  S$+$17  \\
    2M18372953$-$2911046 &      5.94  $\pm$ 0.69 &    36.2   $\pm$ 0.01 &     -1.55 $\pm$ 0.07  &     -2.16  $\pm$ 0.06 &  S$+$17  \\
    2M18442352$-$3029411 &      5.24  $\pm$ 0.43 &    -63.0  $\pm$ 0.25 &     0.36  $\pm$ 0.09  &     -6.26  $\pm$ 0.08 &  S$+$17  \\
    2M18550318$-$3043368 &      14.55 $\pm$ 1.19 &    141.1  $\pm$ 0.14 &     -2.73 $\pm$ 0.08  &     -1.18  $\pm$ 0.07 &  S$+$17  \\
\hline

\end{longtable}
\normalsize 

\clearpage
\onecolumn
\begin{landscape}
\small\addtolength{\tabcolsep}{-4.5pt}
\begin{longtable}{|l | c | c | c | c | c | c | c | c | c | c|}
	\caption{Orbital Elements of selected N$-$rich stars for $\Omega_{\rm bar} =43 $ km s$^{-1}$ kpc$^{-1}$. The number inside parentheses indicate the sensitivity of the orbital parameters to different angular velocity of the bar, which we have computed as the standard deviation of the orbital elements when considering three bar pattern speeds, $\Omega_{\rm bar} = $33, 43, and 53 km s$^{-1}$ kpc$^{-1}$. }
\label{Table2}
\\
\hline
APOGEE\_ID & \texttt{RUWE} &    Z$_{\rm max}$                   &   r$_{\rm peri}$                  &   r$_{\rm apo}$                  &  $ e $                    &            Ej                  &    E$_{\rm char}$                        &    L$_{\rm zmin}$                   &   L$_{\rm zmax}$                   & Orbit        \\
 &  & (kpc) & (kpc) & (kpc)  &  & (10$^{2}$ km$^{2}$ s$^{2}$) & (10$^{2}$ km$^{2}$ s$^{2}$) & (10$^{1}$ km s$^{-1}$ kpc)  & (10$^{1}$ km s$^{-1}$ kpc) &  \\  \endfirsthead
\multicolumn{11}{c}%
{{\bfseries \tablename\ \thetable{} $-$$-$ continued}} \\
\hline
APOGEE\_ID & \texttt{RUWE} &    Z$_{\rm max}$                   &   r$_{\rm peri}$                  &   r$_{\rm apo}$                  &  $ e $                    &            E${\rm j}$                  &    E$_{\rm char}$                        &    L$_{\rm zmin}$                   &   L$_{\rm zmax}$                   & Orbit        \\
 &  & (kpc) & (kpc) & (kpc)  &  & (10$^{2}$ km$^{2}$ s$^{2}$) & (10$^{2}$ km$^{2}$ s$^{2}$) & (10$^{1}$ km s$^{-1}$ kpc)  & (10$^{1}$ km s$^{-1}$ kpc) & \\ 
 \hline
 \endhead

\hline
\endfoot

\hline \hline
\endlastfoot

\hline
\hline
  2M17535944$+$4708092   & 0.95  &   9.70 $\pm$ 0.31 (0.22) &  1.73 $\pm$  0.52 (0.02)  &  15.91 $\pm$ 1.85 (0.59) &  0.80$\pm$ 0.03 (0.01) &  $-$1886.47 $\pm$  48.24 (64.99) &  $-$1542.81$\pm$   46.13 ( 8.16)  &   $-$83.5 $\pm$   22.8 (3.3) &   $-$70.5 $\pm$  19.9 (2.8) &  Prograde            \\
  2M17350460$-$2856477   & 0.96  &   2.12 $\pm$ 0.34 (0.05) &  0.11 $\pm$  0.18 (0.05)  &   3.79 $\pm$ 0.51 (0.13) &  0.94$\pm$ 0.08 (0.02) &  $-$2459.30 $\pm$  34.49 (15.15) &  $-$2354.94$\pm$   87.00 (15.78)  &   $-$34.0 $\pm$   14.8 (2.0) &    $-$1.0 $\pm$   9.4 (1.4) &  Prograde            \\
  2M12155306$+$1431114   & 0.97  &  13.76 $\pm$ 1.03 (0.33) &  3.11 $\pm$  0.39 (0.10)  &  15.65 $\pm$ 0.97 (0.18) &  0.66$\pm$ 0.04 (0.01) &  $-$1932.78 $\pm$  49.44 (81.29) &  $-$1485.90$\pm$   21.15 ( 4.16)  &  $-$104.5 $\pm$    8.4 (3.0) &  $-$102.0 $\pm$   8.9 (3.7) &  Prograde            \\
  2M16062302$-$1126161   & 1.27  &   3.26 $\pm$ 0.23 (0.34) &  1.22 $\pm$  0.34 (0.26)  &   6.22 $\pm$ 0.24 (0.36) &  0.67$\pm$ 0.06 (0.06) &  $-$2282.78 $\pm$  14.39 (38.86) &  $-$2051.79$\pm$   29.10 (34.75)  &   $-$68.0 $\pm$    8.3 (5.8) &   $-$38.0 $\pm$  11.6 (9.6) &  Prograde            \\
  2M17454705$-$2639109   & 0.91  &   0.44 $\pm$ 0.12 (0.02) &  0.21 $\pm$  0.27 (0.06)  &   2.05 $\pm$ 1.21 (0.19) &  0.73$\pm$ 0.15 (0.02) &  $-$2784.80 $\pm$ 179.65 (36.59) &  $-$2669.67$\pm$  260.94 (43.47)  &   $-$31.0 $\pm$   25.3 (4.5) &    $-$5.5 $\pm$   8.4 (1.6) &  Prograde            \\
  2M17492967$-$2328298   & 0.85  &   1.78 $\pm$ 0.32 (0.11) &  0.49 $\pm$  0.18 (0.13)  &   4.02 $\pm$ 0.66 (0.17) &  0.78$\pm$ 0.07 (0.05) &  $-$2489.14 $\pm$  33.19 (27.43) &  $-$2328.99$\pm$   88.42 (21.09)  &   $-$53.0 $\pm$   20.3 (4.2) &   $-$12.0 $\pm$   5.4 (3.7) &  Prograde            \\
  2M17534571$-$2949362   & 0.94  &   0.89 $\pm$ 0.29 (0.10) &  0.39 $\pm$  0.31 (0.09)  &   3.75 $\pm$ 0.83 (0.29) &  0.80$\pm$ 0.09 (0.03) &  $-$2531.23 $\pm$  75.15 (32.91) &  $-$2369.56$\pm$  113.91 (40.68)  &   $-$46.0 $\pm$   16.3 (2.6) &   $-$11.5 $\pm$  12.8 (3.2) &  Prograde            \\
  2M17180311$-$2750124   & 0.86  &   0.45 $\pm$ 0.18 (0.03) &  2.02 $\pm$  0.69 (0.15)  &   5.75 $\pm$ 1.39 (0.55) &  0.49$\pm$ 0.04 (0.05) &  $-$2427.73 $\pm$  14.71 (64.41) &  $-$2103.50$\pm$  108.76 (17.84)  &   $-$92.0 $\pm$   24.9 (6.5) &   $-$63.5 $\pm$  21.9 (5.4) &  Prograde            \\
  2M15204588$+$0055032   & 1.01  &  13.29 $\pm$ 1.21 (0.14) &  0.85 $\pm$  0.81 (0.05)  &  13.27 $\pm$ 1.40 (0.20) &  0.87$\pm$ 0.09 (0.01) &  $-$1762.40 $\pm$  18.08 (23.90) &  $-$1619.55$\pm$  105.31 ( 7.81)  &   $-$38.0 $\pm$   21.5 (0.9) &   $-$28.0 $\pm$  28.4 (1.7) &  Prograde            \\
  2M13251355$-$0044438   & 0.95  &  15.33 $\pm$ 1.47 (0.19) &  0.96 $\pm$  0.08 (0.02)  &  15.48 $\pm$ 1.48 (0.15) &  0.88$\pm$ 0.01 (0.01) &  $-$1638.52 $\pm$  60.13 (24.54) &  $-$1508.70$\pm$   51.70 ( 4.30)  &   $-$33.0 $\pm$    2.4 (0.2) &   $-$30.0 $\pm$   2.4 (0.6) &  Prograde            \\
  2M16514646$-$2127071   & 1.06  &   1.86 $\pm$ 0.39 (0.22) &  0.50 $\pm$  0.24 (0.15)  &   4.48 $\pm$ 0.26 (0.16) &  0.79$\pm$ 0.08 (0.06) &  $-$2428.39 $\pm$  14.88 (18.07) &  $-$2263.01$\pm$   22.64 (25.80)  &   $-$58.0 $\pm$    6.9 (9.7) &   $-$11.5 $\pm$  10.9 (3.5) &  Prograde            \\
  2M17024730$-$2210387   & 0.97  &   1.38 $\pm$ 0.22 (0.10) &  0.73 $\pm$  0.16 (0.09)  &   3.98 $\pm$ 0.46 (0.11) &  0.69$\pm$ 0.05 (0.03) &  $-$2506.36 $\pm$  27.09 (34.11) &  $-$2313.93$\pm$   72.28 (11.06)  &   $-$65.5 $\pm$   16.9 (4.3) &   $-$19.5 $\pm$   4.9 (2.2) &  Prograde            \\
  2M17303980$-$2330234   & 0.99  &   1.16 $\pm$ 0.15 (0.01) &  0.21 $\pm$  0.24 (0.05)  &   1.58 $\pm$ 0.22 (0.04) &  0.74$\pm$ 0.20 (0.07) &  $-$2693.62 $\pm$  26.48 ( 9.72) &  $-$2652.45$\pm$   63.03 ( 2.39)  &   $-$17.0 $\pm$    6.3 (0.2) &    $-$2.5 $\pm$   7.9 (1.3) &  Prograde            \\
  2M17343654$-$1956596   & 1.04  &   1.62 $\pm$ 0.14 (0.05) &  0.64 $\pm$  0.21 (0.23)  &   3.27 $\pm$ 0.67 (0.32) &  0.63$\pm$ 0.07 (0.13) &  $-$2550.09 $\pm$  43.73 (24.96) &  $-$2417.37$\pm$   68.95 (35.52)  &   $-$43.0 $\pm$   13.0 (4.3) &   $-$14.5 $\pm$   5.9 (7.5) &  Prograde            \\
  2M17480576$-$2445000   & 1.27  &   0.31 $\pm$ 0.04 (0.01) &  0.54 $\pm$  0.24 (0.08)  &   2.11 $\pm$ 0.44 (0.04) &  0.60$\pm$ 0.11 (0.05) &  $-$2778.01 $\pm$  73.64 (19.12) &  $-$2664.14$\pm$   97.82 ( 9.38)  &   $-$28.5 $\pm$   10.8 (3.2) &   $-$12.5 $\pm$   6.4 (2.0) &  Prograde            \\
  2M17482995$-$2305299   & 0.99  &   1.17 $\pm$ 0.24 (0.19) &  0.02 $\pm$  0.01 (0.01)  &   2.50 $\pm$ 0.57 (0.03) &  0.98$\pm$ 0.01 (0.01) &  $-$2615.43 $\pm$  64.94 ( 8.15) &  $-$2585.73$\pm$   98.30 (16.34)  &   $-$18.0 $\pm$    7.9 (3.0) &    11.5 $\pm$   5.4 (2.0) &  Prograde            \\
  2M17494963$-$2318560   & 0.99  &   0.46 $\pm$ 0.04 (0.01) &  0.23 $\pm$  0.21 (0.05)  &   1.72 $\pm$ 0.22 (0.03) &  0.76$\pm$ 0.16 (0.05) &  $-$2827.42 $\pm$  34.77 (13.52) &  $-$2730.11$\pm$   59.70 ( 6.11)  &   $-$23.0 $\pm$    3.4 (1.2) &    $-$4.5 $\pm$   4.9 (1.2) &  Prograde            \\
  2M17504980$-$2255083   & 0.94  &   1.64 $\pm$ 0.22 (0.04) &  1.45 $\pm$  0.52 (0.37)  &   3.45 $\pm$ 0.51 (0.16) &  0.40$\pm$ 0.09 (0.06) &  $-$2487.09 $\pm$  41.74 (54.17) &  $-$2305.00$\pm$   99.78 (31.21)  &   $-$65.0 $\pm$   12.6 (3.2) &   $-$30.0 $\pm$  13.1 (8.5) &  Prograde            \\
  2M17554454$-$2123058   & 1.13  &   1.25 $\pm$ 0.61 (0.40) &  0.38 $\pm$  0.30 (0.04)  &   4.45 $\pm$ 0.88 (0.04) &  0.84$\pm$ 0.09 (0.01) &  $-$2423.19 $\pm$  61.55 (15.11) &  $-$2288.12$\pm$  113.74 ( 7.14)  &   $-$41.5 $\pm$   15.4 (2.3) &   $-$12.0 $\pm$  10.4 (1.2) &  Prograde            \\
  2M18022530$-$2928338   & 0.90  &   0.67 $\pm$ 0.37 (0.04) &  0.89 $\pm$  0.56 (0.02)  &   4.84 $\pm$ 1.08 (0.02) &  0.70$\pm$ 0.14 (0.01) &  $-$2466.75 $\pm$  71.64 (46.16) &  $-$2228.02$\pm$  126.99 (10.04)  &   $-$76.5 $\pm$   19.9 (6.2) &   $-$28.0 $\pm$  19.4 (1.0) &  Prograde            \\
  2M18032356$-$3001588   & 0.83  &   1.27 $\pm$ 0.10 (0.02) &  0.68 $\pm$  0.15 (0.04)  &   2.86 $\pm$ 0.48 (0.09) &  0.60$\pm$ 0.03 (0.01) &  $-$2601.25 $\pm$  38.95 (24.51) &  $-$2477.06$\pm$   57.21 ( 6.76)  &   $-$42.5 $\pm$    9.0 (1.0) &   $-$16.0 $\pm$   4.9 (1.8) &  Prograde            \\
  2M18120031$-$1350169   & 1.10  &   1.22 $\pm$ 0.25 (0.04) &  0.14 $\pm$  0.11 (0.08)  &   3.09 $\pm$ 0.29 (0.27) &  0.91$\pm$ 0.06 (0.03) &  $-$2576.72 $\pm$  27.68 (16.23) &  $-$2457.87$\pm$   58.55 (28.35)  &   $-$38.0 $\pm$    6.4 (7.8) &    $-$2.5 $\pm$   6.4 (1.4) &  Prograde            \\
  2M18124455$-$2719146   & 1.09  &   1.40 $\pm$ 0.28 (0.10) &  0.18 $\pm$  0.13 (0.04)  &   2.77 $\pm$ 0.86 (0.16) &  0.89$\pm$ 0.06 (0.01) &  $-$2596.34 $\pm$  99.66 ( 8.02) &  $-$2499.41$\pm$  142.01 (13.58)  &   $-$32.0 $\pm$   11.3 (4.1) &    $-$3.0 $\pm$   4.4 (0.8) &  Prograde            \\
  2M18165340$-$2017051   & 1.05  &   0.39 $\pm$ 0.08 (0.04) &  0.45 $\pm$  0.15 (0.03)  &   2.98 $\pm$ 0.66 (0.20) &  0.75$\pm$ 0.05 (0.01) &  $-$2642.19 $\pm$  86.52 (37.08) &  $-$2478.22$\pm$  113.35 (44.18)  &   $-$49.0 $\pm$    7.4 (4.5) &   $-$12.0 $\pm$   5.4 (1.0) &  Prograde            \\
  2M18372953$-$2911046   & 0.91  &   1.17 $\pm$ 0.09 (0.01) &  1.10 $\pm$  0.33 (0.11)  &   2.86 $\pm$ 0.22 (0.17) &  0.44$\pm$ 0.05 (0.05) &  $-$2583.72 $\pm$  19.93 (40.69) &  $-$2438.37$\pm$   51.22 (20.98)  &   $-$48.0 $\pm$    9.4 (3.3) &   $-$24.5 $\pm$   7.8 (4.0) &  Prograde            \\
  2M18442352$-$3029411   & 1.00  &   1.29 $\pm$ 0.12 (0.14) &  0.56 $\pm$  0.20 (0.01)  &   4.32 $\pm$ 0.38 (0.22) &  0.76$\pm$ 0.05 (0.01) &  $-$2491.14 $\pm$  16.40 (27.13) &  $-$2273.41$\pm$   58.32 (30.82)  &   $-$75.5 $\pm$   19.9 (1.7) &   $-$15.5 $\pm$   7.4 (0.4) &  Prograde            \\
  2M18550318$-$3043368   & 0.93  &  10.56 $\pm$ 1.89 (0.14) &  1.29 $\pm$  0.14 (0.02)  &  10.70 $\pm$ 1.72 (0.22) &  0.77$\pm$ 0.02 (0.01) &  $-$1504.91 $\pm$  96.20 (42.53) &  $-$1692.94$\pm$   89.05 (14.44)  &    39.0 $\pm$    4.4 (0.6) &    44.5 $\pm$   3.5 (0.6) &  Retrograde          \\
  2M02491285$+$5534213   & 0.96  &   8.09 $\pm$ 0.21 (0.50) &  1.57 $\pm$  0.24 (0.05)  &   9.33 $\pm$ 0.04 (0.21) &  0.71$\pm$ 0.04 (0.01) &  $-$1571.11 $\pm$  38.70 (41.08) &  $-$1810.41$\pm$   15.60 (17.91)  &    49.0 $\pm$    7.9 (1.6) &    60.0 $\pm$   8.9 (3.2) &  Retrograde          \\
  2M15113526$+$3551140   & 1.05  &  15.07 $\pm$ 1.20 (0.35) &  0.90 $\pm$  0.09 (0.09)  &  15.35 $\pm$ 1.24 (0.46) &  0.89$\pm$ 0.01 (0.01) &  $-$1386.27 $\pm$  70.90 (29.32) &  $-$1542.19$\pm$   51.49 (10.04)  &    33.0 $\pm$    7.3 (4.7) &    42.0 $\pm$   7.4 (3.4) &  Retrograde          \\
  2M17252263$+$4903137   & 0.97  &  12.55 $\pm$ 2.01 (0.13) &  0.24 $\pm$  0.12 (0.05)  &  16.14 $\pm$ 1.45 (0.61) &  0.96$\pm$ 0.01 (0.01) &  $-$1517.27 $\pm$  44.78 ( 8.27) &  $-$1580.37$\pm$   45.87 (22.10)  &     6.5 $\pm$    4.9 (1.6) &    21.0 $\pm$   5.9 (1.8) &  Retrograde          \\
  2M16011638$-$1201525   & 1.38  &   7.64 $\pm$ 2.47 (0.17) &  3.04 $\pm$  0.21 (0.01)  &  11.87 $\pm$ 2.59 (0.07) &  0.58$\pm$ 0.08 (0.01) &  $-$1186.54 $\pm$ 103.09 (89.31) &  $-$1659.06$\pm$   97.37 ( 0.79)  &   107.0 $\pm$    1.5 (0.4) &   111.0 $\pm$   1.5 (0.4) &  Retrograde          \\
  2M16493657$-$2028146   & 1.05  &   2.49 $\pm$ 0.34 (0.01) &  0.02 $\pm$  0.01 (0.01)  &   2.54 $\pm$ 0.60 (0.06) &  0.97$\pm$ 0.01 (0.01) &  $-$2473.46 $\pm$  49.79 ( 3.49) &  $-$2441.47$\pm$   52.31 (20.44)  &   $-$15.0 $\pm$    8.9 (3.1) &     5.5 $\pm$   6.9 (8.0) &  Prograde/Retrograde \\
  2M17134700$-$2441353   & 0.99  &   1.83 $\pm$ 0.19 (0.03) &  0.04 $\pm$  0.03 (0.01)  &   2.17 $\pm$ 0.71 (0.06) &  0.95$\pm$ 0.03 (0.01) &  $-$2563.98 $\pm$  38.86 ( 7.87) &  $-$2614.60$\pm$   78.72 ( 8.28)  &    $-$5.5 $\pm$    7.9 (1.0) &    21.0 $\pm$   6.4 (0.6) &  Prograde/Retrograde \\
  2M17161691$-$2458586   & 0.91  &   1.37 $\pm$ 0.10 (0.03) &  0.10 $\pm$  0.31 (0.03)  &   2.28 $\pm$ 1.10 (0.04) &  0.89$\pm$ 0.15 (0.03) &  $-$2647.35 $\pm$  76.11 (17.34) &  $-$2588.12$\pm$  173.90 ( 2.36)  &   $-$25.0 $\pm$   25.3 (1.8) &     0.0 $\pm$  14.9 (3.7) &  Prograde/Retrograde \\
  2M17173203$-$2439094   & 0.98  &   1.25 $\pm$ 0.30 (0.07) &  0.04 $\pm$  0.05 (0.01)  &   1.84 $\pm$ 0.85 (0.01) &  0.96$\pm$ 0.08 (0.01) &  $-$2637.37 $\pm$  71.19 ( 8.60) &  $-$2640.61$\pm$  114.88 (16.22)  &   $-$11.0 $\pm$    8.9 (3.3) &    13.0 $\pm$   2.9 (1.2) &  Prograde/Retrograde \\
  2M17193271$-$2732214   & 0.85  &   1.54 $\pm$ 0.14 (0.04) &  0.01 $\pm$  0.01 (0.01)  &   2.05 $\pm$ 0.23 (0.01) &  0.98$\pm$ 0.01 (0.01) &  $-$2647.51 $\pm$  22.36 ( 6.74) &  $-$2650.44$\pm$   37.93 (11.25)  &    $-$8.0 $\pm$    2.5 (1.6) &    13.5 $\pm$   3.4 (0.4) &  Prograde/Retrograde \\
  2M17205201$-$2903061   & 0.97  &   0.70 $\pm$ 0.06 (0.02) &  0.02 $\pm$  0.03 (0.01)  &   1.78 $\pm$ 0.69 (0.08) &  0.96$\pm$ 0.02 (0.01) &  $-$2770.87 $\pm$  90.55 (13.77) &  $-$2720.00$\pm$  147.48 (21.38)  &   $-$20.0 $\pm$   14.3 (2.2) &     3.5 $\pm$   4.4 (1.8) &  Prograde/Retrograde \\
  2M17211817$-$2735530   & 1.12  &   0.95 $\pm$ 0.13 (0.03) &  0.02 $\pm$  0.01 (0.01)  &   1.54 $\pm$ 0.72 (0.12) &  0.96$\pm$ 0.02 (0.01) &  $-$2754.33 $\pm$  79.65 (13.52) &  $-$2746.83$\pm$  123.62 (19.05)  &   $-$11.5 $\pm$   13.4 (1.8) &     5.0 $\pm$   0.5 (0.9) &  Prograde/Retrograde \\
  2M17263951$-$2406247   & 1.06  &   0.85 $\pm$ 0.08 (0.01) &  0.04 $\pm$  0.04 (0.01)  &   1.94 $\pm$ 0.82 (0.19) &  0.95$\pm$ 0.03 (0.01) &  $-$2733.08 $\pm$  83.80 (23.05) &  $-$2668.85$\pm$  135.31 (26.17)  &   $-$25.5 $\pm$   14.9 (4.1) &     1.0 $\pm$   2.9 (0.4) &  Prograde/Retrograde \\
  2M17271907$-$2718040   & 1.12  &   2.43 $\pm$ 0.40 (0.25) &  0.04 $\pm$  0.05 (0.01)  &   4.04 $\pm$ 0.50 (0.12) &  0.97$\pm$ 0.02 (8.88) &  $-$2387.12 $\pm$  28.47 ( 6.64) &  $-$2357.11$\pm$   65.04 (17.09)  &   $-$24.5 $\pm$   12.4 (5.5) &    11.0 $\pm$   8.9 (3.1) &  Prograde/Retrograde \\
  2M17305251$-$2651528   & 1.01  &   1.20 $\pm$ 0.18 (0.03) &  0.03 $\pm$  0.03 (0.01)  &   2.24 $\pm$ 0.80 (0.05) &  0.96$\pm$ 0.02 (0.01) &  $-$2651.95 $\pm$  77.52 ( 8.83) &  $-$2587.92$\pm$  104.28 ( 9.87)  &   $-$24.0 $\pm$    9.0 (0.4) &     0.5 $\pm$   2.9 (0.2) &  Prograde/Retrograde \\
  2M17334208$-$2958347   & 1.09  &   0.31 $\pm$ 0.08 (0.01) &  0.01 $\pm$  0.01 (0.01)  &   1.15 $\pm$ 0.44 (0.10) &  0.97$\pm$ 0.01 (8.70) &  $-$2913.25 $\pm$  73.89 (23.85) &  $-$2897.85$\pm$  101.00 (25.69)  &   $-$10.5 $\pm$    9.4 (1.7) &     6.0 $\pm$   1.9 (0.8) &  Prograde/Retrograde \\
  2M17343610$-$2909472   & 1.32  &   0.66 $\pm$ 0.35 (0.01) &  0.02 $\pm$  0.01 (0.01)  &   1.58 $\pm$ 0.64 (0.23) &  0.96$\pm$ 0.02 (0.01) &  $-$2810.21 $\pm$ 105.32 (32.60) &  $-$2753.63$\pm$  141.90 (52.45)  &   $-$19.0 $\pm$   11.9 (5.2) &     2.5 $\pm$   3.0 (0.2) &  Prograde/Retrograde \\
  2M17343807$-$2557555   & 0.99  &   0.55 $\pm$ 0.11 (0.05) &  0.02 $\pm$  0.02 (0.01)  &   1.98 $\pm$ 0.62 (0.08) &  0.97$\pm$ 0.01 (0.01) &  $-$2751.68 $\pm$ 103.89 ( 7.71) &  $-$2680.50$\pm$  153.97 (14.17)  &   $-$23.0 $\pm$   10.9 (3.2) &     2.0 $\pm$   2.0 (0.1) &  Prograde/Retrograde \\
  2M17352288$-$2913255   & 1.13  &   0.33 $\pm$ 0.08 (0.04) &  0.02 $\pm$  0.15 (0.01)  &   2.26 $\pm$ 0.78 (0.05) &  0.97$\pm$ 0.08 (0.01) &  $-$2721.82 $\pm$ 106.93 (13.32) &  $-$2623.09$\pm$  152.77 ( 8.73)  &   $-$29.5 $\pm$   12.4 (3.4) &     1.0 $\pm$   6.9 (0.4) &  Prograde/Retrograde \\
  2M17353215$-$2759106   & 1.06  &   0.40 $\pm$ 0.03 (0.01) &  0.02 $\pm$  0.05 (0.01)  &   1.02 $\pm$ 0.46 (0.13) &  0.95$\pm$ 0.05 (0.01) &  $-$2935.03 $\pm$  70.85 (22.04) &  $-$2925.87$\pm$  107.77 (39.75)  &    $-$9.5 $\pm$    9.3 (2.7) &     2.0 $\pm$   1.5 (1.6) &  Prograde/Retrograde \\
  2M17354267$-$2406233   & 0.73  &   1.28 $\pm$ 0.17 (0.07) &  0.01 $\pm$  0.01 (0.01)  &   2.32 $\pm$ 0.06 (0.02) &  0.98$\pm$ 0.01 (0.01) &  $-$2646.69 $\pm$  18.39 (14.74) &  $-$2603.15$\pm$   23.55 (20.24)  &   $-$20.0 $\pm$    3.4 (3.3) &     6.5 $\pm$   2.9 (0.8) &  Prograde/Retrograde \\
  2M17382269$-$2748001   & 0.95  &   0.38 $\pm$ 0.06 (0.01) &  0.01 $\pm$  0.01 (0.01)  &   1.10 $\pm$ 0.34 (0.28) &  0.97$\pm$ 0.01 (0.01) &  $-$2925.30 $\pm$  55.17 (62.86) &  $-$2908.22$\pm$   68.23 (71.04)  &   $-$10.0 $\pm$    4.9 (4.8) &     5.0 $\pm$   1.0 (0.6) &  Prograde/Retrograde \\
  2M17390422$-$2943520   & 0.87  &   1.44 $\pm$ 0.49 (0.02) &  0.02 $\pm$  0.03 (0.01)  &   1.83 $\pm$ 1.17 (0.10) &  0.96$\pm$ 0.02 (0.01) &  $-$2645.49 $\pm$ 137.56 (19.03) &  $-$2615.07$\pm$  187.46 (42.80)  &   $-$18.5 $\pm$   11.4 (4.5) &     6.0 $\pm$   5.9 (1.3) &  Prograde/Retrograde \\
  2M17404143$-$2714570   & 0.95  &   0.86 $\pm$ 0.54 (0.17) &  0.02 $\pm$  0.02 (0.01)  &   1.86 $\pm$ 0.63 (0.01) &  0.97$\pm$ 0.01 (0.01) &  $-$2688.89 $\pm$ 101.09 (12.52) &  $-$2722.30$\pm$  109.75 ( 5.31)  &    $-$9.5 $\pm$    7.0 (3.1) &    15.0 $\pm$   3.0 (0.9) &  Prograde/Retrograde \\
  2M17434675$-$2616068   & 1.23  &   2.27 $\pm$ 0.64 (0.21) &  0.04 $\pm$  0.02 (0.01)  &   3.53 $\pm$ 0.78 (0.33) &  0.97$\pm$ 0.01 (0.01) &  $-$2470.68 $\pm$ 111.04 (31.82) &  $-$2407.12$\pm$  112.92 (40.53)  &   $-$21.5 $\pm$    4.9 (0.9) &     6.5 $\pm$   6.5 (1.1) &  Prograde/Retrograde \\
  2M17442343$-$2627304   & 0.94  &   0.56 $\pm$ 0.12 (0.07) &  0.01 $\pm$  0.01 (0.01)  &   1.57 $\pm$ 0.44 (0.05) &  0.97$\pm$ 0.01 (0.01) &  $-$2816.42 $\pm$  74.79 (17.35) &  $-$2792.80$\pm$  103.94 ( 7.94)  &   $-$15.0 $\pm$    6.9 (3.3) &     9.0 $\pm$   2.0 (0.4) &  Prograde/Retrograde \\
  2M17453131$-$2342147   & 0.92  &   0.63 $\pm$ 0.12 (0.01) &  0.02 $\pm$  0.05 (0.02)  &   1.52 $\pm$ 0.09 (0.07) &  0.96$\pm$ 0.06 (0.03) &  $-$2819.81 $\pm$  37.54 (17.71) &  $-$2767.80$\pm$   33.17 (17.50)  &   $-$17.0 $\pm$    3.9 (0.6) &     2.5 $\pm$   3.9 (2.4) &  Prograde/Retrograde \\
  2M17464449$-$2531533   & 0.96  &   0.39 $\pm$ 0.11 (0.01) &  0.01 $\pm$  0.01 (0.01)  &   1.17 $\pm$ 0.41 (0.15) &  0.97$\pm$ 0.01 (0.01) &  $-$2910.24 $\pm$  62.11 (16.15) &  $-$2896.68$\pm$   77.23 (24.71)  &   $-$10.5 $\pm$    7.4 (3.7) &     6.0 $\pm$   1.9 (0.4) &  Prograde/Retrograde \\
  2M17523300$-$3027521   & 1.26  &   1.99 $\pm$ 0.61 (0.01) &  0.14 $\pm$  0.23 (0.14)  &   3.64 $\pm$ 1.06 (0.15) &  0.91$\pm$ 0.09 (0.04) &  $-$2459.07 $\pm$  86.98 (11.52) &  $-$2407.38$\pm$  151.42 (32.20)  &   $-$27.0 $\pm$   27.8 (8.6) &     4.5 $\pm$  20.3 (6.9) &  Prograde/Retrograde \\
  2M18014817$-$3026237   & 1.07  &   0.60 $\pm$ 0.01 (0.01) &  0.01 $\pm$  0.01 (0.01)  &   0.51 $\pm$ 0.21 (0.04) &  0.96$\pm$ 0.01 (0.01) &  $-$2945.66 $\pm$  15.83 ( 2.92) &  $-$2945.28$\pm$   23.04 ( 4.79)  &    $-$1.5 $\pm$    2.4 (0.6) &     1.5 $\pm$   0.9 (0.4) &  Prograde/Retrograde \\
  2M18033335$-$2929122   & 1.07  &   0.56 $\pm$ 0.10 (0.01) &  0.15 $\pm$  0.14 (0.05)  &   0.95 $\pm$ 0.72 (0.02) &  0.71$\pm$ 0.06 (0.09) &  $-$2867.13 $\pm$  74.77 (27.40) &  $-$2866.03$\pm$  143.63 ( 8.75)  &    $-$4.5 $\pm$   16.7 (1.0) &     0.5 $\pm$   7.9 (0.2) &  Prograde/Retrograde \\
  2M18035944$-$2908195   & 0.77  &   1.48 $\pm$ 0.32 (0.17) &  0.04 $\pm$  0.02 (0.01)  &   3.12 $\pm$ 0.14 (0.04) &  0.97$\pm$ 0.01 (0.01) &  $-$2537.38 $\pm$  15.65 ( 4.44) &  $-$2477.83$\pm$   31.20 (10.03)  &   $-$23.5 $\pm$    6.3 (3.4) &     2.5 $\pm$   3.9 (2.5) &  Prograde/Retrograde \\
  2M18054875$-$3122407   & 1.01  &   0.80 $\pm$ 0.09 (0.01) &  0.01 $\pm$  0.04 (0.01)  &   0.83 $\pm$ 0.38 (0.03) &  0.96$\pm$ 0.05 (0.02) &  $-$2847.34 $\pm$  46.71 ( 5.37) &  $-$2855.47$\pm$   74.01 ( 8.42)  &    $-$4.5 $\pm$    8.9 (0.7) &     2.0 $\pm$   3.5 (0.8) &  Prograde/Retrograde \\
  2M18061336$-$3147053   & 0.90  &   0.83 $\pm$ 0.09 (0.01) &  0.02 $\pm$  0.03 (0.01)  &   1.75 $\pm$ 0.66 (0.21) &  0.96$\pm$ 0.02 (0.01) &  $-$2758.17 $\pm$  75.45 (29.16) &  $-$2702.73$\pm$  115.09 (43.83)  &   $-$19.5 $\pm$   11.9 (4.7) &     2.0 $\pm$   2.0 (0.8) &  Prograde/Retrograde \\
  2M18090957$-$1559276   & 0.93  &   0.64 $\pm$ 0.52 (0.07) &  0.03 $\pm$  0.21 (0.01)  &   2.68 $\pm$ 0.39 (0.12) &  0.97$\pm$ 0.13 (0.01) &  $-$2655.95 $\pm$  49.77 (14.41) &  $-$2540.13$\pm$   71.90 (11.54)  &   $-$34.5 $\pm$   16.3 (3.3) &     0.0 $\pm$  14.3 (4.9) &  Prograde/Retrograde \\
  2M18102953$-$2707208   & 0.97  &   1.40 $\pm$ 0.16 (0.13) &  0.01 $\pm$  0.01 (0.01)  &   1.96 $\pm$ 0.69 (0.23) &  0.98$\pm$ 0.01 (0.01) &  $-$2648.74 $\pm$  83.08 (19.05) &  $-$2649.94$\pm$   90.25 (38.57)  &    $-$8.5 $\pm$    4.9 (5.4) &    13.0 $\pm$   1.9 (0.2) &  Prograde/Retrograde \\
  2M18121957$-$2926310   & 0.92  &   0.88 $\pm$ 0.19 (0.04) &  0.01 $\pm$  0.07 (0.01)  &   1.05 $\pm$ 0.40 (0.05) &  0.96$\pm$ 0.08 (0.01) &  $-$2823.13 $\pm$  74.51 (10.34) &  $-$2801.98$\pm$   75.14 ( 2.30)  &    $-$7.0 $\pm$    3.5 (0.4) &     4.0 $\pm$   4.4 (1.4) &  Prograde/Retrograde \\  
  2M18334592$-$2903253   & 1.00  &   1.98 $\pm$ 0.20 (0.09) &  0.03 $\pm$  0.10 (0.07)  &   3.26 $\pm$ 0.55 (0.16) &  0.97$\pm$ 0.04 (0.03) &  $-$2491.59 $\pm$  44.76 ( 6.43) &  $-$2469.61$\pm$  109.55 (41.14)  &   $-$26.5 $\pm$   12.6 (5.9) &    10.5 $\pm$  10.9 (6.0) &  Prograde/Retrograde \\
\end{longtable}
\end{landscape}

\normalsize
\twocolumn 

\onecolumn

\scriptsize\addtolength{\tabcolsep}{-0.5pt}
\begin{longtable}{|l |c c c| c c c |c c c|}
\caption{Membership probability for different bar pattern speed.}
\label{Table3}
\\
\hline
 & \multicolumn{3}{c|}{$\Omega_{\rm bar}=33$ $\mathrm{km} \,\mathrm{s}^{-1}\, \mathrm{kpc}^{-1}$} & \multicolumn{3}{c|}{$\Omega_{\rm bar}=43$ $\mathrm{km} \,\mathrm{s}^{-1}\, \mathrm{kpc}^{-1}$}&\multicolumn{3}{c|}{$\Omega_{\rm bar}=53$ $\mathrm{km} \,\mathrm{s}^{-1}\, \mathrm{kpc}^{-1}$} \\
 \hline
APOGEE\_ID &  Bulge/Bar & Disk & Stellar Halo  &  Bulge/Bar & Disk & Stellar Halo & Bulge/Bar & Disk & Stellar Halo \\
 &\multicolumn{3}{c|}{\%} & \multicolumn{3}{c|}{\%}  &\multicolumn{3}{c|}{\%} \\  \endfirsthead

\multicolumn{10}{c}%
{{\bfseries \tablename\ \thetable{} -- continued}} \\
\hline
APOGEE\_ID &  Bulge/Bar & Disk & Stellar Halo  &  Bulge/Bar & Disk & Stellar Halo & Bulge/Bar & Disk & Stellar Halo \\
 &\multicolumn{3}{c|}{\%} & \multicolumn{3}{c|}{\%}  &\multicolumn{3}{c|}{\%} \\ 
 \hline
 \endhead

\hline
\endfoot

\hline \hline
\endlastfoot

\hline
\hline 
   2M17535944$+$4708092 &  0.0        &  0.0       & 100.0     &  0.0        &  0.06      &  99.94     &  0.0         &  36.42     &  63.58     \\
   2M17350460$-$2856477 &  22.34      &  76.64     & 1.02      &  11.82      &  87.45     &  0.73      &  7.92        &  91.04     &  1.04      \\
   2M12155306$+$1431114 &  0.0        &  0.0       & 100.0     &  0.0        &  0.34      &  99.66     &  0.0         &  39.13     &  60.87     \\
   2M16062302$-$1126161 &  0.59       &  95.51     & 3.9       &  0.33       &  95.02     &  4.65      &  0.26        &  96.51     &  3.23      \\
   2M17454705$-$2639109 &  99.96      &  0.0       & 0.04      &  99.03      &  0.94      &  0.03      &  82.66       &  17.27     &  0.07      \\
   2M17492967$-$2328298 &  47.8       &  51.13     & 1.07      &  12.44      &  86.94     &  0.62      &  16.49       &  82.86     &  0.65      \\
   2M17534571$-$2949362 &  98.01      &  1.66      & 0.33      &  23.13      &  76.37     &  0.49      &  43.3        &  56.38     &  0.32      \\
   2M17180311$-$2750124 &  0.76       &  98.34     & 0.9       &  1.3        &  97.79     &  0.91      &  1.42        &  97.75     &  0.83      \\
   2M02491285$+$5534213 &  0.0        &  0.0       & 100.0     &  0.0        &  0.0       &  100.0     &  0.0         &  0.0       &  100.0     \\
   2M15113526$+$3551140 &  0.0        &  0.0       & 100.0     &  0.0        &  0.0       &  100.0     &  0.0         &  0.0       &  100.0     \\
   2M15204588$+$0055032 &  0.0        &  0.0       & 100.0     &  0.0        &  0.0       &  100.0     &  0.0         &  1.52      &  98.48     \\
   2M13251355$-$0044438 &  0.0        &  0.0       & 100.0     &  0.0        &  0.0       &  100.0     &  0.0         &  0.16      &  99.84     \\
   2M17252263$+$4903137 &  0.0        &  0.0       & 100.0     &  0.0        &  0.0       &  100.0     &  0.0         &  0.0       &  100.0     \\
   2M16011638$-$1201525 &  0.0        &  0.0       & 100.0     &  0.0        &  0.0       &  100.0     &  0.0         &  0.0       &  100.0     \\
   2M16493657$-$2028146 &  64.43      &  34.87     & 0.69      &  23.75      &  75.65     &  0.6       &  19.65       &  79.4      &  0.95      \\
   2M16514646$-$2127071 &  11.96      &  87.08     & 0.95      &  4.62       &  94.44     &  0.93      &  5.5         &  93.28     &  1.22      \\
   2M17024730$-$2210387 &  34.68      &  64.19     & 1.14      &  12.9       &  86.53     &  0.58      &  14.15       &  85.22     &  0.63      \\
   2M17134700$-$2441353 &  99.01      &  0.86      & 0.13      &  76.71      &  23.14     &  0.16      &  58.38       &  41.31     &  0.31      \\
   2M17161691$-$2458586 &  99.85      &  0.07      & 0.08      &  87.76      &  12.14     &  0.1       &  75.93       &  23.96     &  0.11      \\
   2M17173203$-$2439094 &  99.95      &  0.02      & 0.04      &  90.75      &  9.18      &  0.07      &  80.25       &  19.65     &  0.1       \\
   2M17193271$-$2732214 &  99.92      &  0.03      & 0.05      &  92.38      &  7.55      &  0.06      &  78.19       &  21.7      &  0.11      \\
   2M17205201$-$2903061 &  99.98      &  0.0       & 0.02      &  99.21      &  0.77      &  0.02      &  90.47       &  9.49      &  0.04      \\
   2M17211817$-$2735530 &  99.99      &  0.0       & 0.01      &  99.19      &  0.79      &  0.02      &  96.44       &  3.55      &  0.01      \\
   2M17263951$-$2406247 &  99.99      &  0.0       & 0.01      &  97.89      &  2.08      &  0.03      &  94.03       &  5.95      &  0.02      \\
   2M17271907$-$2718040 &  18.01      &  81.07     & 0.91      &  7.12       &  91.55     &  1.33      &  5.86        &  92.39     &  1.75      \\
   2M17303980$-$2330234 &  99.97      &  0.0       & 0.03      &  95.82      &  4.14      &  0.05      &  86.05       &  13.89     &  0.06      \\
   2M17305251$-$2651528 &  99.87      &  0.05      & 0.08      &  88.37      &  11.53     &  0.1       &  71.81       &  28.06     &  0.13      \\
   2M17334208$-$2958347 &  100.0      &  0.0       & 0.0       &  99.98      &  0.01      &  0.0       &  97.94       &  2.05      &  0.01      \\
   2M17343610$-$2909472 &  100.0      &  0.0       & 0.0       &  99.68      &  0.31      &  0.01      &  98.3        &  1.69      &  0.01      \\
   2M17343654$-$1956596 &  97.99      &  1.71      & 0.3       &  35.3       &  64.29     &  0.41      &  46.59       &  53.11     &  0.3       \\
   2M17343807$-$2557555 &  99.98      &  0.0       & 0.02      &  98.54      &  1.43      &  0.03      &  87.91       &  12.05     &  0.05      \\
   2M17352288$-$2913255 &  99.96      &  0.0       & 0.04      &  96.39      &  3.56      &  0.05      &  82.34       &  17.58     &  0.07      \\
   2M17353215$-$2759106 &  100.0      &  0.0       & 0.0       &  99.99      &  0.01      &  0.0       &  98.02       &  1.97      &  0.01      \\
   2M17354267$-$2406233 &  99.81      &  0.09      & 0.09      &  88.97      &  10.94     &  0.09      &  71.41       &  28.45     &  0.14      \\
   2M17382269$-$2748001 &  99.99      &  0.0       & 0.01      &  99.99      &  0.01      &  0.0       &  95.2        &  4.78      &  0.02      \\
   2M17390422$-$2943520 &  99.98      &  0.0       & 0.02      &  89.77      &  10.14     &  0.08      &  88.51       &  11.43     &  0.05      \\
   2M17404143$-$2714570 &  99.99      &  0.0       & 0.01      &  97.48      &  2.49      &  0.03      &  91.1        &  8.86      &  0.04      \\
   2M17434675$-$2616068 &  95.39      &  4.27      & 0.34      &  18.59      &  80.77     &  0.64      &  36.18       &  63.31     &  0.51      \\
   2M17442343$-$2627304 &  99.99      &  0.0       & 0.01      &  99.79      &  0.2       &  0.01      &  95.81       &  4.18      &  0.02      \\
   2M17453131$-$2342147 &  99.99      &  0.0       & 0.01      &  99.76      &  0.23      &  0.01      &  94.89       &  5.09      &  0.02      \\
   2M17464449$-$2531533 &  100.0      &  0.0       & 0.0       &  99.98      &  0.02      &  0.0       &  99.18       &  0.82      &  0.0       \\
   2M17480576$-$2445000 &  99.99      &  0.0       & 0.01      &  98.88      &  1.1       &  0.03      &  92.65       &  7.32      &  0.03      \\
   2M17482995$-$2305299 &  99.69      &  0.19      & 0.12      &  82.47      &  17.4      &  0.13      &  59.12       &  40.64     &  0.24      \\
   2M17494963$-$2318560 &  99.99      &  0.0       & 0.01      &  99.7       &  0.28      &  0.01      &  95.47       &  4.51      &  0.02      \\
   2M17504980$-$2255083 &  5.71       &  93.34     & 0.96      &  10.26      &  89.1      &  0.64      &  5.76        &  93.54     &  0.7       \\
   2M17523300$-$3027521 &  26.8       &  72.23     & 0.97      &  17.11      &  82.19     &  0.7       &  6.88        &  91.75     &  1.37      \\
   2M17554454$-$2123058 &  9.8        &  89.24     & 0.96      &  5.4        &  93.63     &  0.97      &  4.21        &  94.48     &  1.31      \\
   2M18014817$-$3026237 &  100.0      &  0.0       & 0.0       &  99.99      &  0.01      &  0.0       &  99.1        &  0.9       &  0.0       \\
   2M18022530$-$2928338 &  3.73       &  95.35     & 0.91      &  4.8        &  94.47     &  0.72      &  3.71        &  95.36     &  0.93      \\
   2M18032356$-$3001588 &  97.14      &  2.49      & 0.37      &  61.54      &  38.2      &  0.26      &  45.53       &  54.18     &  0.29      \\
   2M18033335$-$2929122 &  100.0      &  0.0       & 0.0       &  99.95      &  0.04      &  0.0       &  98.63       &  1.36      &  0.01      \\
   2M18035944$-$2908195 &  91.66      &  7.85      & 0.49      &  44.06      &  55.57     &  0.37      &  30.9        &  68.57     &  0.53      \\
   2M18054875$-$3122407 &  100.0      &  0.0       & 0.0       &  99.93      &  0.07      &  0.01      &  98.08       &  1.91      &  0.01      \\
   2M18061336$-$3147053 &  99.96      &  0.0       & 0.04      &  98.9       &  1.08      &  0.02      &  81.41       &  18.51     &  0.08      \\
   2M18090957$-$1559276 &  99.82      &  0.09      & 0.09      &  84.29      &  15.57     &  0.13      &  71.09       &  28.77     &  0.14      \\
   2M18102953$-$2707208 &  99.85      &  0.08      & 0.07      &  92.47      &  7.47      &  0.06      &  68.07       &  31.75     &  0.19      \\
   2M18120031$-$1350169 &  98.59      &  1.15      & 0.26      &  50.63      &  49.05     &  0.32      &  48.44       &  51.27     &  0.29      \\
   2M18121957$-$2926310 &  100.0      &  0.0       & 0.0       &  99.83      &  0.16      &  0.01      &  97.08       &  2.91      &  0.01      \\
   2M18124455$-$2719146 &  98.76      &  0.99      & 0.26      &  64.52      &  35.24     &  0.24      &  46.43       &  53.26     &  0.31      \\
   2M18165340$-$2017051 &  99.94      &  0.0       & 0.05      &  72.82      &  26.96     &  0.21      &  80.25       &  19.67     &  0.08      \\
   2M18334592$-$2903253 &  54.21      &  44.73     & 1.06      &  31.77      &  67.73     &  0.5       &  12.41       &  86.74     &  0.84      \\
   2M18372953$-$2911046 &  81.57      &  17.57     & 0.86      &  48.58      &  51.08     &  0.33      &  31.12       &  68.51     &  0.37      \\
   2M18442352$-$3029411 &  34.84      &  64.0      & 1.16      &  8.39       &  90.97     &  0.63      &  12.15       &  87.15     &  0.7       \\
   2M18550318$-$3043368 &  0.0        &  0.0       & 100.0     &  0.0        &  0.0       &  100.0     &  0.0         &  0.0       &  100.0     \\
\end{longtable} 
 
\normalsize
\twocolumn

\label{lastpage}
\end{document}